\documentclass[12pt]{article}
\usepackage{times}
\usepackage{amsmath,amssymb}
\usepackage{epsfig,graphics}
\usepackage{epstopdf}

\usepackage[makeroom]{cancel}
\usepackage{color}

\begin{document}
\hsize = 6. in
\vsize = 9.0 in
\hoffset = -0.5 in
\voffset = -0.85 in
\baselineskip = 0.23 in
\def\rd{{\rm d}}
\def\wtu{\widetilde u}
\def\vz{{\bf z}}
\def\vdelta{\mbox{\boldmath$\delta$}}
\def\mD{{\bf D}}
\def\mI{{\bf I}}
\def\mJ{{\bf J}}
\def\vf{{\bf f}}
\def\mF{{\bf F}}
\def\rvx{{\bf x}}
\def\rvy{{\bf y}}

\title{A Thermodynamic Theory of Ecology: Helmholtz Theorem for
Lotka-Volterra Equation, Extended Conservation
Law, and Stochastic Predator-Prey Dynamics}

\author{Yi-An Ma\footnote{Email: yianma@u.washington.edu }\hspace{0.2cm}
  and Hong Qian\footnote{
Email: hqian@u.washington.edu}\\[10pt]
Department of Applied Mathematics\\
University of Washington, Seattle\\
WA 98195-3925, U.S.A.}

\maketitle

%
%
%


\begin{abstract}
We carry out mathematical analyses, {\em \`{a} la} Helmholtz's and Boltzmann's 1884 studies of monocyclic Newtonian dynamics, for the Lotka-Volterra (LV) equation exhibiting predator-prey oscillations.  In doing so a novel ``thermodynamic theory" of ecology is introduced. 
An important feature, absent in the classical mechanics, of ecological systems is a natural stochastic population dynamic formulation of which the deterministic equation (e.g., the LV equation studied) is the infinite population limit.  Invariant density for the stochastic dynamics plays a central role in the deterministic LV dynamics. We show how the conservation law along a single trajectory extends to incorporate both variations in a model parameter $\alpha$ and in initial conditions: Helmholtz's theorem establishes a broadly valid conservation law in a class of ecological dynamics.  We analyze the relationships among mean ecological activeness $\theta$, quantities characterizing dynamic ranges of populations $\mathcal{A}$ and $\alpha$, and the ecological force $F_{\alpha}$. The analyses identify an entire orbit as a stationary ecology, and establish the notion of ``equation of ecological states". Studies of the stochastic dynamics with finite populations show the LV equation as the robust, fast cyclic underlying behavior. The mathematical narrative provides a novel way of capturing long-term dynamical behaviors with an emergent {\em conservative ecology}.
\end{abstract}

\section{Introduction}

	In the mathematical investigations of ecological systems, conservative
dynamics are often considered non-robust, thus unrealistic as a
faithful description of reality \cite{jdmurray,mkot}.   Through recent
studies of stochastic, nonlinear population dynamics, however, a
new perspective has emerged \cite{qian_iop,zqq,aoping}:  The stationary
behavior of a complex stochastic population dynamics almost always
exhibits a divergence-free cyclic motion in its phase space,
even when the corresponding system of differential equations has
only stable, node-type fixed points \cite{qqw_tpb}.   In particular, it has been
shown that an underlying volume preserving conservative dynamics is one of the
essential keys to understand the long-time complexity of such stochastic
systems \cite{wangjin,qian_jmp,qian_pla}.

The aim of the present work, following a proposal in
\cite{qian_pla}, is to carry out a comprehesive stochastic dynamic
and thermodynamic analysis of an ecological system with
sustained oscillations.  In the
classical studies on statistical mechanics, developed by Helmholtz,
Boltzmann, Gibbs, and others, the dynamical foundation is a
Hamiltonian system \cite{gallavotti,khinchin,mcampisi_05}.
The theory in \cite{qian_pla,qian_ma_linear} generalized such an approach that
requires no {\em a priori} identification of variables
as position and momentum; it also suggested a possible {\em
thermodynamic structure} which is purely mathematical in
nature, independent of Newtonian particles.  In the context of
population dynamics, we shall show that the mathematical analysis
yields a {\em conservative ecology}.

	Among ecological models, the Lotka-Volterra (LV) equation for
predator-prey system has played an important pedagogical role \cite{jdmurray,mkot,lotka},
even though it is certainly not a realistic model for any engineering
applications.  We choose this population system in the
present work because its mathematics tractability, and its stochastic
counterpart in terms of a birth and death process \cite{linda_allen,grasman}.
It can be rigorously shown that a smooth solusion to LV differential equation is the
law of large numbers for the stochastic process \cite{kurtz}.  In
biochemistry, the birth-death process for discrete,
stochastic reactions corresponding to the mass-action kinetics has
been called a Delbr\"{u}ck-Gillespie process \cite{qian_iop}.

	The LV equation in non-dimensionalized form is \cite{jdmurray}:
\begin{equation}
       \frac{\rd x}{\rd t} = x(1-y) = f(x,y), \  \
	\frac{\rd y}{\rd t} = \alpha y(x-1) = g(x,y;\alpha),
\label{theeqn}
\end{equation}
in which $x(t)$ and $y(t)$ represent the populations of a
prey and its predator, each normalized with respect to its
time-averaged mean populations.  The $xy$ term in $f(x,y)$
stands for the rate of consumption of the prey by the
predator, and the $\alpha y x$ term in $g(x,y;\alpha)$ stands
for the rate of prey-dependent predator growth.

\begin{figure}[h]
\begin{center}
\includegraphics[width=2.8in]{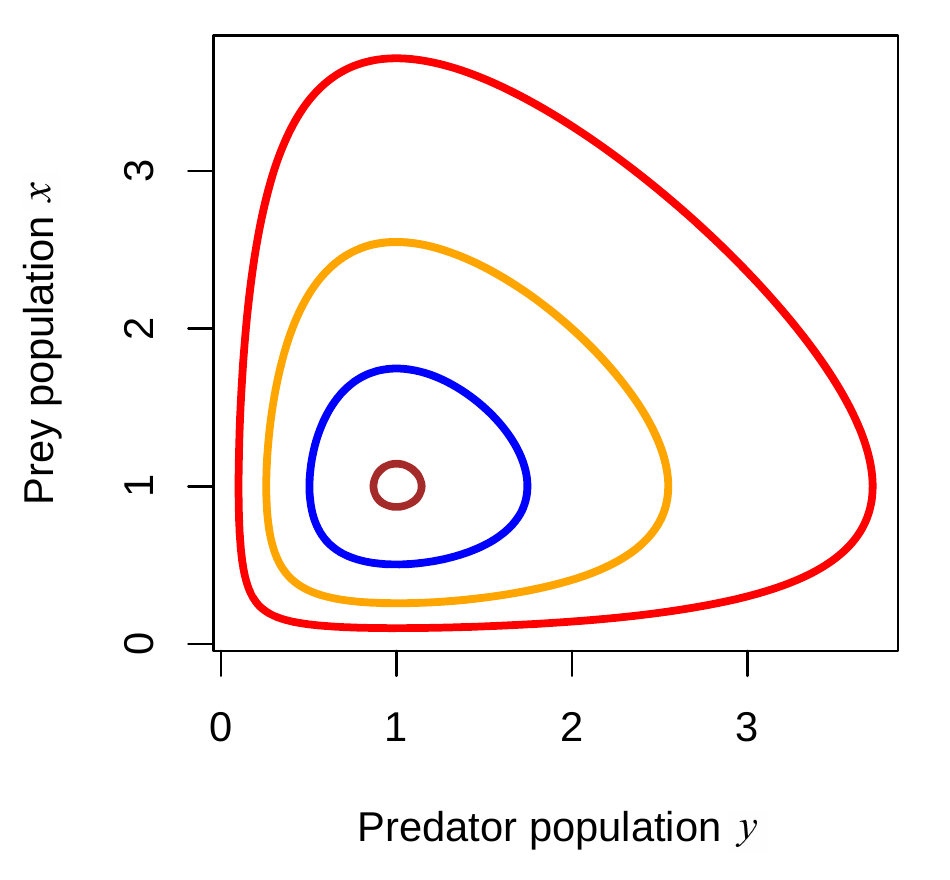}
\includegraphics[width=2.8in]{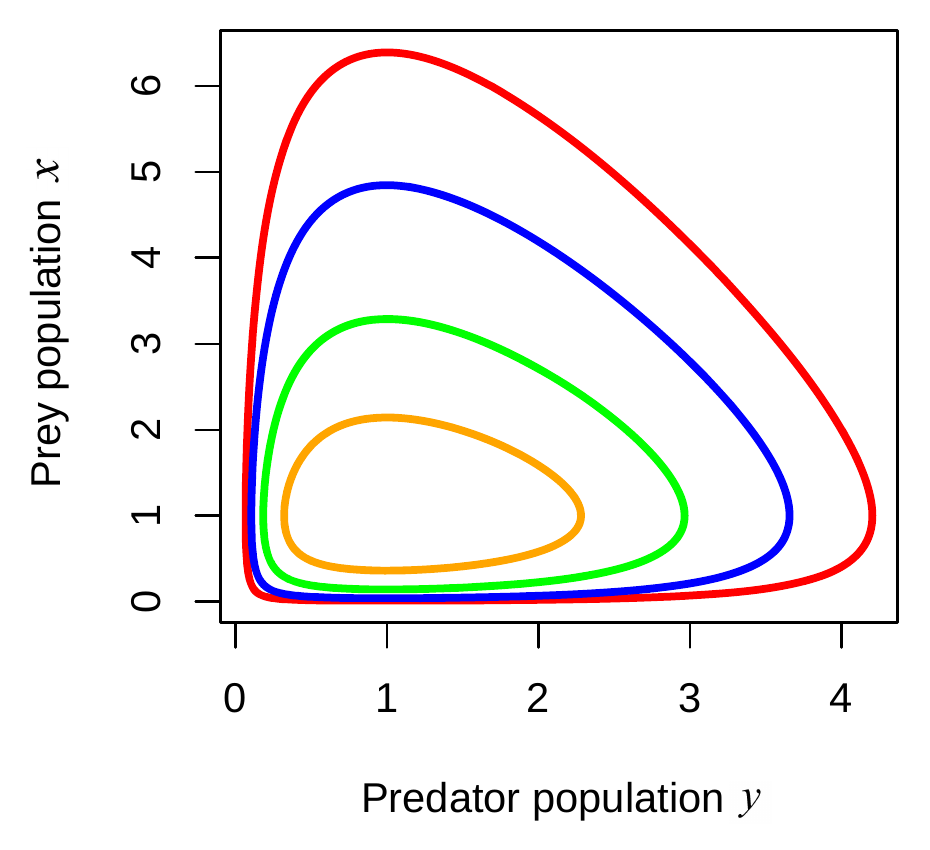}
\end{center}
\caption{Left panel: with $\alpha=1$ and $H(x,y)=3.40,2.61,2.19$,
and $2.01$. Right panel:  with $\alpha = 0.5, 0.6, 0.8$, and $1.2$,
from outside inward, all with $H(x,y)=2.61$.  We see that the larger
the $\alpha$, the smaller the temporal variations in the prey
population, relative to  that of predator.
}
\label{fig_1}
\end{figure}

It is easy to check that the solutions to (\ref{theeqn}) in
phase space are level curves of a scalar
function \cite{jdmurray}
\begin{equation}
          H(x,y) = \alpha x + y - \ln \big( x^{\alpha} y\big).
\label{Hxy}
\end{equation}
We shall use $\Gamma_{H=h}$ to denote the solution
curve $H(x,y)=h$, and $\mathfrak{D}_{h}(\alpha)$ to denote the
domain encircled by the $\Gamma_{H=h}$.  Fig. \ref{fig_1} shows
the contours of $H(x,y)$ with $\alpha=1$ and
$H(x,y)=2.61$ with different $\alpha$'s.

Let $\tau$ be the period of the cyclic dynamics.  Then it is easy to
show that
\cite{jdmurray}
\begin{equation}
    \frac{1}{\tau}\int_0^{\tau} x(t) \rd t = \frac{1}{\tau}\int_0^{\tau} y(t) \rd t  = 1.
\end{equation}
Furthermore (see Appendix \ref{app-A}),
\begin{eqnarray}
    && \frac{1}{\tau}\int_0^{\tau} \big(x(t)-1\big)^2 \rd t = \frac{\hat{\mathcal{A}}}{\alpha\tau},
\\
	&& \frac{1}{\tau}\int_0^{\tau} \big(y(t)-1\big)^2 \rd t = \frac{\alpha\hat{\mathcal{A}}}{\tau},
\end{eqnarray}
in which $\hat{\mathcal{A}}$ is the area of $\mathfrak{D}_h(\alpha)$,
encircled by $\Gamma_{H=h}$, using Lebesgue measure in the $xy$-plane.
The appropriate measure for computing
the area will be further discussed in Sec. \ref{sec2}.
The parameter $\alpha$ represents the relative temporal variations,
or dynamic ranges, in
the two populations: the larger the $\alpha$, the greater the temporal
variations and range in the predator population, and the smaller in
the prey population.

The paper is organized as follows. In Sec. \ref{sec2},
an extended conservation law is recognized for the Lotka-Volterra system.
Then the relationship among three quantities: the ``energy" function $H(x,y)$,
the area $\mathcal{A}$ encircled by the level set $\Gamma_{H=h}$,
and the parameter $\alpha$ is developed.  According to the Helmholtz
theorem, the conjugate variables of $\mathcal{A}$ and $\alpha$ are found
as time averages of certain functions of population $x(t)$ and $y(t)$.
Analysis on those novel ``state variables" demonstrates
the tendency of change in mean ecological quantities like population range or ecological activeness when the parameter $\alpha$ or energy $H$ varies.
In Sec. \ref{sec3}, we show that the area $\mathcal{A}$ encircled by $\Gamma_{H=h}$ is related to the concept of entropy.  In Sec. \ref{sec4},
the conservative dynamics is shown to be an integral part of the stochastic population dynamics, which necessarily has the same invariant density
as the deterministic conservative dynamics.  In the large population limit, a separation of time scale between the fast cyclic motion on $\Gamma_{H=h}$ and the slow stochastic crossing of $\Gamma_{H=h}$ is observed in the stochastic dynamical system.  The paper concludes with a discussion in
Sec. \ref{sec5}.

\section{The Helmholtz theorem}
\label{sec2}

Eq. (\ref{theeqn})  is not a Hamiltonian system, nor is it divergence-free
\[
    \frac{\partial f(x,y)}{\partial x} + \frac{\partial g(x,y;\alpha)}{\partial y}
         \neq 0.
\]
It can be expressed, however, as
\begin{equation}
        \frac{\rd x}{\rd t} = -G(x,y)\frac{\partial H(x,y)}{\partial y}, \  \  \
        \frac{\rd y}{\rd t} =  G(x,y)\frac{\partial H(x,y)}{\partial x},
\label{non-df}
\end{equation}
with a scalar factor $G(x,y)=xy$.  One can in fact
understand this scalar factor as a ``local change of measure'',
or time $\rd\hat{t}\equiv G\big(x(t),y(t)\big) \rd t$ \cite{qian_jmp}:
\begin{eqnarray}
      && x(t) = \hat{x}\big(\hat{t}(t)\big),  \  \
            y(t) = \hat{y}\big(\hat{t}(t)\big),
\end{eqnarray}
for
\begin{eqnarray}
\label{t2that}
      && \hat{t}(t) = \hat{t}_0 + \int_{t_0}^t
                 G\big(x(s),y(s)\big)  \rd s,
\nonumber
\end{eqnarray}
where $(\hat{x},\hat{y})$ satisfies the corresponding
Hamiltonian system.   In Sec. \ref{sec3} and \ref{sec4} below,
we shall show that $G^{-1}(x,y)$ is an invariant density
of the Liouville equation for the deterministic dynamics (\ref{theeqn}),
and more importantly the invariant density of the Fokker-Planck
equation for the corresponding stochastic dynamics.
As will be demonstrated in Sec. \ref{sec2_2} and \ref{sec2_3}, statistical average of quantities according to the invariant measure $G^{-1}(x,y)\rd x \rd y$ can be calculated through time average of those quantities along the system's instantaneous dynamics.
Knowledge about the system's long term distribution is not needed during the calculation.
These facts make the $G^{-1}(x,y)$ the natural measure for computing
area $\mathcal{A}$.

Any function of $H(x,y)$, $\rho(H)$ is conserved under
the dynamics,
as is guaranteed by the orthogonality between the
vector field of (\ref{theeqn}) and gradient $\nabla\rho$ \cite{qian_pla}:
\begin{eqnarray}
    \frac{\rd \rho\big(H(x,y)\big)} {\rd t}
    &=& f(x,y)\frac{\partial}{\partial x} \rho\big(H(x,y)\big) +
    g(x,y;\alpha) \frac{\partial}{\partial y} \rho\big(H(x,y)\big)
\nonumber\\
	&=& \rho'(H)\left( x(1-y)\alpha\left(1-\frac{1}{x}\right)
                         + \alpha y(x-1)\left(1-\frac{1}{y}\right) \right)
\nonumber\\
	&=& 0.
\label{invmeasure}
\end{eqnarray}
This is analogous to the ``conservation law" observed in Hamiltonian systems.

\subsection{Extending the conservation law}

	The nonlinear dynamics in (\ref{theeqn}), therefore,
introduces a ``conservative relation'' between the populations
of predator and prey according to (\ref{Hxy}).  If we
call the  value $H(x,y)$ an ``energy'',
then the phase portrait in the left panel of Fig. \ref{fig_1}
suggests that the entire phase space of the dynamical system is
organized according to the value of $H$.  The deep
insight contained in the work of Helmholtz and Boltzmann
\cite{gallavotti,mcampisi_05} is that
such an energy-based organization can be further extended
for different values of $\alpha$:  Therefore, the
energy-based organization is no longer limited to a
{\em single} orbit, nor a {\em single} dynamical system; but rather
for the entire class of parametric dynamical systems.  In the
classical physics of Newtonian mechanical energy
conservation, this yields the mechanical basis of the
Fundamental Thermodynamic Relation as a form of
the First Law, which extends the
notion of energy conservation far beyond mechanical
systems \cite{planck,pauli}.

	More specifically, we see that the area $\mathcal{A}$ in
Fig. \ref{fig_1}, or in fact any geometric quantification of
a closed orbit, is completely determined by the parameter
$\alpha$ and initial energy value $h=H\big(x(0),y(0),\alpha\big)$.
Therefore, there must exist a bivariate function
$ \mathcal{A} = \mathcal{A}(h,\alpha)$,
Assuming the implicit function theorem applies, then one
has
\begin{equation}
          h =  h(\mathcal{A},\alpha).
\label{hfunction}
\end{equation}
Note that in terms of the Eq. (\ref{hfunction}), a ``state''
of the ecological system is not a single point $(x,y)$ which
is continuously varying with time; rather it reflects the geometry of an entire orbit.
Then Eq. (\ref{hfunction}) implies that any such ecological
state has an ``h-energy'', {\em if} one recognizes a
geometric, state variable $\mathcal{A}$.

	Eq. (\ref{hfunction}) can be written in a differential
form
\begin{equation}
            \rd h = \left(\frac{\partial h}{\partial \mathcal{A}}\right)_{\alpha}
                  \rd \mathcal{A} +\left(\frac{\partial h}{\partial \alpha}\right)_{\mathcal{A}}
                  \rd \alpha,
\label{dh}
\end{equation}
in which one first introduces the h-energy for an ecological system
with fixed $\alpha$ via the factor $(\partial h/\partial\mathcal{A})$.
Then, holding $\mathcal{A}$ constant, one introduces an
``$\alpha$-force'' corresponding to the
parameter $\alpha$.
In classical thermodynamics,
the latter is known as an ``adiabatic'' process.

	The Helmholtz theorem expresses the two
partial derivatives in (\ref{dh}) in terms of the dynamics in
Eq. (\ref{theeqn}).

\subsection{Projected invariant measure}
\label{sec2_2}
For canonical Hamiltonian systems, Lebesgue measure is an invariant
measure in the whole phase space.
On the level set $\Gamma_{H=h}$, the projection of the Lebesgue measure,
called the Liouville measure, also defines an invariant measure on
the sub-manifold.
If the dynamics on the invariant sub-manifold $\Gamma_{H=h}$ is ergodic,
the average with respect to the Liouville measure is equal to the time average
along the trajectory starting from any initial condition $(x_0,y_0)$
satisfying $H(x_0,y_0)=h$.

	As we shall show below, the invariant measure for
the LV system (\ref{theeqn}) in the whole phase space is
$\rd \mathcal{A} = G^{-1}(x,y) \rd x\rd y$.
Projection of this invariant measure onto the level set $\Gamma_{H=h}$
can be found by changing $(x,y)$ to intrinsic coordinates $(h,\ell)$:
\begin{equation}
\rd \mathcal{A} = G^{-1}(x,y) \ \rd x\rd y
= G^{-1}(x,y) \ \big(\rd x, \rd y\big)^T \cdot \mathbf{n} \ \rd \ell,
\end{equation}
where
\begin{eqnarray}
\mathbf{n} = \left(\dfrac{\partial H(x,y) / \partial x}{||\nabla H(x,y)||} , \dfrac{\partial H(x,y) / \partial y}{||\nabla H(x,y)||}\right)_{(x,y)\in\Gamma_{H=h}}^T
\end{eqnarray}
is the unit normal vector of the the level set $\Gamma_{H=h}$;
and $\rd \ell = \sqrt{\rd x^2 + \rd y^2}$.
Noting that:
\begin{eqnarray}
\rd h &=& \dfrac{\partial H(x,y)}{\partial x} \rd x + \dfrac{\partial H(x,y)}{\partial y} \rd y, \end{eqnarray}
we have
\begin{eqnarray}
\big(\rd x, \rd y\big)^T \cdot \mathbf{n} = \dfrac{\rd h}{||\nabla H(x,y)||}.
\end{eqnarray}
That is:
\begin{equation}
           \rd \mathcal{A} = \rd \mu \ \rd h,
\end{equation}
where
\begin{equation}
\rd \mu = \dfrac{G^{-1}(x,y)}{||\nabla H(x,y)||} \rd \ell
\end{equation}
is the projected invariant measure of the Lotka-Volterra system on the level set $\Gamma_{H=h}$.

It is worth noting that $\rd \mu=\rd t$ on the level set $\Gamma_{H=h}$.
Since dynamics on $\Gamma_{H=h}$ is ergodic, the average of any function $\psi(x,y)$ under the projected invariant measure on $\Gamma_{H=h}$ is equal to its time average over a period:
\begin{align}
              \langle \psi\rangle^{\Gamma_{H=h}}
              &\triangleq \frac{\displaystyle
                     \oint_{\Gamma_{H=h}} \psi\big(x,y\big)  \rd \mu}
                {\displaystyle  \oint_{\Gamma_{H=h}}  \rd \mu} \nonumber\\
                     &=\frac{1}{\tau}
                     \int_0^{\tau} \psi\big(x(t),y(t)\big) \rd t
                     \triangleq \langle \psi\rangle^t
\end{align}

\subsection{Functional relation between
\boldmath{$\ln\mathcal{A}$}, \boldmath{$\alpha$}, and \boldmath{$h$} }
\label{sec2_3}
Under the invariant measure $G^{-1}(x,y)$,
the area $\mathcal{A}$ encircled by the level curve $\Gamma_{H=h}$ is:
\begin{eqnarray}
                 \mathcal{A}_{\mathfrak{D}_h(\alpha)} &=&
               \iint_{\mathfrak{D}_h(\alpha)}   G^{-1}(x,y)
                                 \rd x\rd y
                                 \nonumber\\
                                 &=&
                   \iint_{\mathfrak{D}_h(\alpha)}
                                 \rd\ln x\ \rd\ln y
\end{eqnarray}
Using Green's theorem the area
$\mathcal{A}_{\mathfrak{D}_h(\alpha)}$ can be simplified as
\begin{eqnarray}
\mathcal{A}_{\mathfrak{D}_h(\alpha)}
	&=& \int_0^{\tau(h,\alpha)}
	\ln y\left(\frac{\partial H}{\partial \ln y}\right) \rd t
\nonumber\\
    &=& \int_0^{\tau(h,\alpha)}
    \ln x\left(\frac{\partial H}{\partial \ln x}\right) \rd t,
\end{eqnarray}
where $\tau(h,\alpha)$ is the time period for the cyclic motion.
Furthermore,
\begin{eqnarray}
	  \frac{\partial \mathcal{A}_{\mathfrak{D}_h(\alpha)}}
                 {\partial h} &=&
      \frac{\partial}{\partial h} \iint_{\mathfrak{D}_h(\alpha)}
      G^{-1}(x,y) \ \rd x \rd y
\nonumber\\
      &=& \int_0^{\tau(h,\alpha)} \rd t = \tau(h,\alpha).
\end{eqnarray}
That is
\begin{equation}
           \left( \frac{\partial\ln\mathcal{A}_{\mathfrak{D}_h(\alpha)}} {\partial h}
           \right)^{-1}   =   \left\langle  \ln x \left(\frac{\partial H}{\partial\ln x}\right)
                                  \right\rangle^{t}
                =  \left\langle \ln y \left(\frac{\partial H}{\partial\ln y}\right)\right\rangle^{t},
\label{eq0014}
\end{equation}
in which $\langle \cdots \rangle^{t}$ is the time average, or phase space average according to the invariant measure.
We can also find the derivative of the area $\mathcal{A}_{\mathfrak{D}_h(\alpha)}$ encircled by the level curve $\Gamma_{H=h}$ with respect to the parameter of the system $\alpha$ as:
\begin{eqnarray}
	  \frac{\partial \mathcal{A}_{\mathfrak{D}_h(\alpha)}}
                 {\partial \alpha} &=&
      \frac{\partial}{\partial \alpha} \iint_{\mathfrak{D}_h(\alpha)}
      G^{-1}(x,y) \  \rd x \rd y
\nonumber\\
	&=&  - \int_0^{\tau(h,\alpha)}  \big(x(t)-\ln{x(t)}\big) \ \rd t.
\end{eqnarray}

In this setting, the Helmholtz theorem reads
\begin{equation}
        \rd h  =  \frac{\displaystyle \rd\mathcal{A}
                  - \left(\frac{\partial\mathcal{A}}{\partial\alpha}
                    \right)_{h} \rd\alpha } {\displaystyle
                       \left(\frac{\partial\mathcal{A}}{\partial h}
                    \right)_{\alpha} } =  \theta(h,\alpha)\
                \rd\ln\mathcal{A}
                  - F_{\alpha}(h,\alpha) \rd\alpha,
\label{dheq28}
\end{equation}
in which
\begin{equation}
          \theta(h,\alpha)
          = \mathcal{A}_{\mathfrak{D}_h(\alpha)} \left(\frac{\partial\mathcal{A}}{\partial h}
                    \right)_{\alpha} ^{-1}
          = \left\langle \ln x \left(\frac{\partial H}{\partial\ln x}\right) \right\rangle^{t}
          = \left\langle \ln y \left(\frac{\partial H}{\partial\ln y}\right) \right\rangle^{t}.
\label{temp}
\end{equation}
The factor $\theta(h,\alpha)$  here is the mean
$\ln x (\partial H/\partial\ln x)$, or $\ln y (\partial H/\partial\ln y)$,
precisely like the mean kinetic energy as the notion of temperature in
classical physics, and the virial theorem.
The $\alpha$-force is then defined as
\begin{equation}
     F_{\alpha}(h,\alpha) =  \left(\frac{\partial \mathcal{A}}{\partial\alpha}
                    \right)_{h}  \left(\frac{\partial\mathcal{A}}{\partial h}
                    \right)_{\alpha} ^{-1}
                    = - \left\langle \frac{\partial H(x,y,\alpha)}
                  {\partial \alpha} \right\rangle^{t}.
\label{F-alpha}
\end{equation}

It is important to note that the definition of $F_{\alpha}$
given in the right-hand-side of (\ref{F-alpha}) is completely
independent of the notion of $\mathcal{A}$, even though the
relation (\ref{dheq28}) explicitly involves the latter.
$F_{\alpha}(h,\alpha)$ is a function of both $h$ and $\alpha$,
however.  Therefore, the value of $\alpha$-work
$F_{\alpha}(h,\alpha)\rd\alpha$ depends on how $h$
is constrained: There are iso-$h$ processes, iso-$\theta$ processes,
etc. \cite{qian-epjst}

\subsection{Equation of state}

The notion of an equation of state first appeared in
classic thermodynamics \cite{planck,pauli}.   From a modern
dynamical systems standpoint,  a fixed point
as a function usually is continuously dependent
upon the parameters in a mathematical model, except
at bifurcation points.  Let $(x_1^*,x_2^*,\cdots,
x_n^*)$ be a globally asymptotically attractive fixed point,
and $\alpha$ be a parameter, then the function $x_1^*(\alpha)$
constitutes an {\em equation of state} for
the long-time ``equilibrium'' behavior of the dynamical system.

	If a system has a globally asymptotically attractive
limit set that is not a simple fixed point, then
every geometric characteristic of the invariant
manifold, say $\mathfrak{g}^*$, will be a function of $\alpha$.
In this case, $\mathfrak{g}^*(\alpha)$ could well be considered
as an equation of state.  An ``equilibrium state'' in this case
is the entire invariant manifold.

	The situation for a conservative dynamical system with
center manifolds is quite different.  In this
case, the long-time behavior of the dynamical system,
the foremost, is dependent upon its initial data.   An
equation of state therefore is a functional relation
among (i) geometric characteristics of a center manifold
$\mathfrak{g}^*$, (ii) parameter $\alpha$,
and (iii) a new quantity, or quantities, that identifies a
specific center manifold, $h$.   This is the fundamental insight
of the Helmholtz theorem.

In ecological terms, area under the invariant measure: $\mathcal{A}$,
gives a sense of total variation in both the predator's and the prey's populations.
Therefore, $\ln\mathcal{A}$ measures population range of both populations as a whole.
The parameter $\alpha$, on the other hand, is the proportion of predators' over preys' population ranges of time variations:
\begin{eqnarray}
\alpha^2=\dfrac{\int_0^{\tau(h,\alpha)} \big(y(t)-1\big)^2 \rd t}{\int_0^{\tau(h,\alpha)} \big(x(t)-1\big)^2 \rd t}
= \dfrac{\left\langle (y-1)^2 \right\rangle^t}{\left\langle (x-1)^2 \right\rangle^t}.
\label{alpha}
\end{eqnarray}
The new quantity $\theta$ can be
viewed as a measure of the mean ecological ``activeness'':
\begin{eqnarray}
\theta=\left\langle \alpha(x-1)\ln x \right\rangle^t
=\left\langle (y-1)\ln y \right\rangle^t.
\label{eq2700}
\end{eqnarray}
It is the mean of ``distance" from the prey's and predator's populations $x$ and $y$, to the fixed populations in equilibrium $(1,1)$.  For
population dynamic variable $u$,  Eq. \ref{eq2700} suggests a norm
$\| u\|\equiv u\ln (u+1)$.  Then, $\theta=\left\langle \alpha \|x-1\| \right\rangle^t=\left\langle \|y-1\| \right\rangle^t$;
and an averaged norm of per capita growth rates in the two species:
\begin{eqnarray}
\theta
=\left\langle \alpha \left\|\frac{1}{\alpha}\dfrac{\rd \ln y}{\rd t}\right\| \right\rangle^t
=\left\langle  \left\|-\dfrac{\rd \ln x}{\rd t}\right\| \right\rangle^t.
\end{eqnarray}
And finally,
\begin{equation}
             F_\alpha = - \left\langle \dfrac{\partial H(x,y,\alpha)}{\partial \alpha} \right\rangle^{t}
             = - \left\langle x - \ln x \right\rangle^{t}
             \label{F_alpha}
\end{equation}
is the ``ecological force" one needs to counteract in order to
change $\alpha$.
In other words, when $|F_\alpha|$ is greater, more $h$-energy change is needed to vary $\alpha$.
It is also worth noting that $|F_\alpha|$ is positively related to the prey's average population range.
In fact we can define another ``distance" of the prey's population $x$ to $1$ as: $\|u\|_F=u-\ln (u+1)$,
then $F_\alpha=-\big\langle \|x-1\|_F \big\rangle^{t} - 1 $.  Note that for
very small $u$:  $\|u\|\approx u^2\approx 2\|u\|_F$


\begin{figure}
\begin{center}
\includegraphics[width=1.9in]{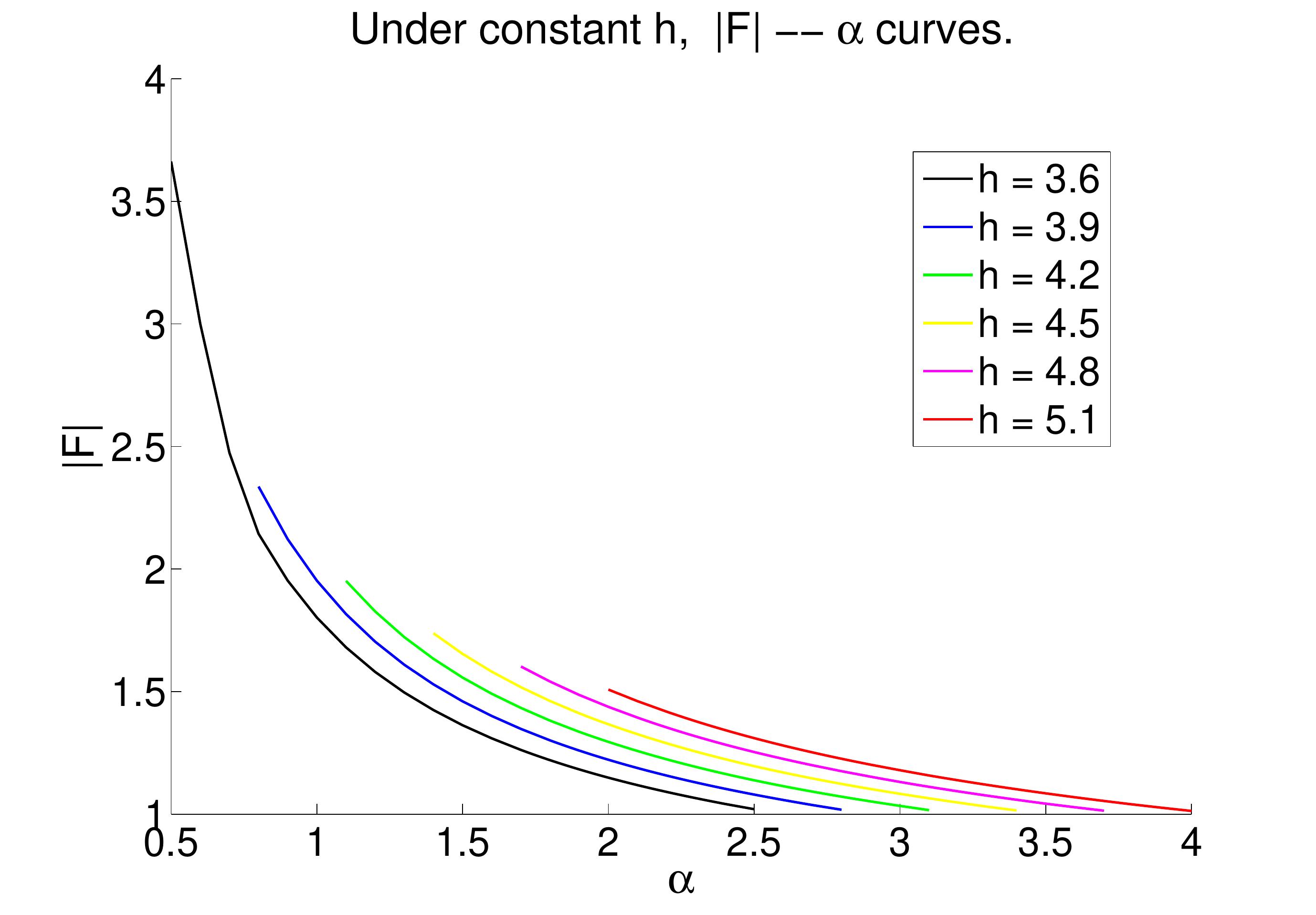}
\includegraphics[width=1.9in]{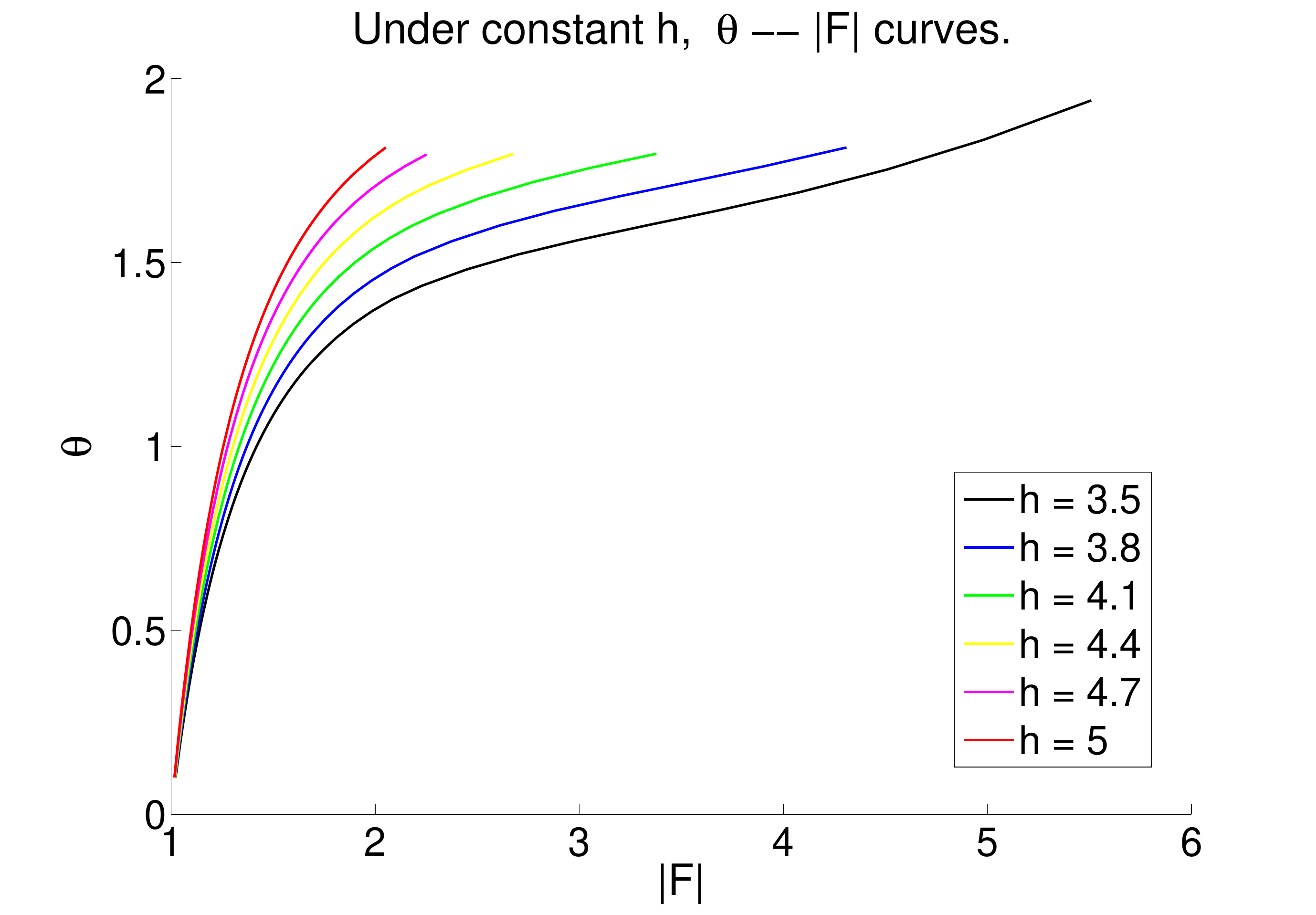}
\includegraphics[width=1.9in]{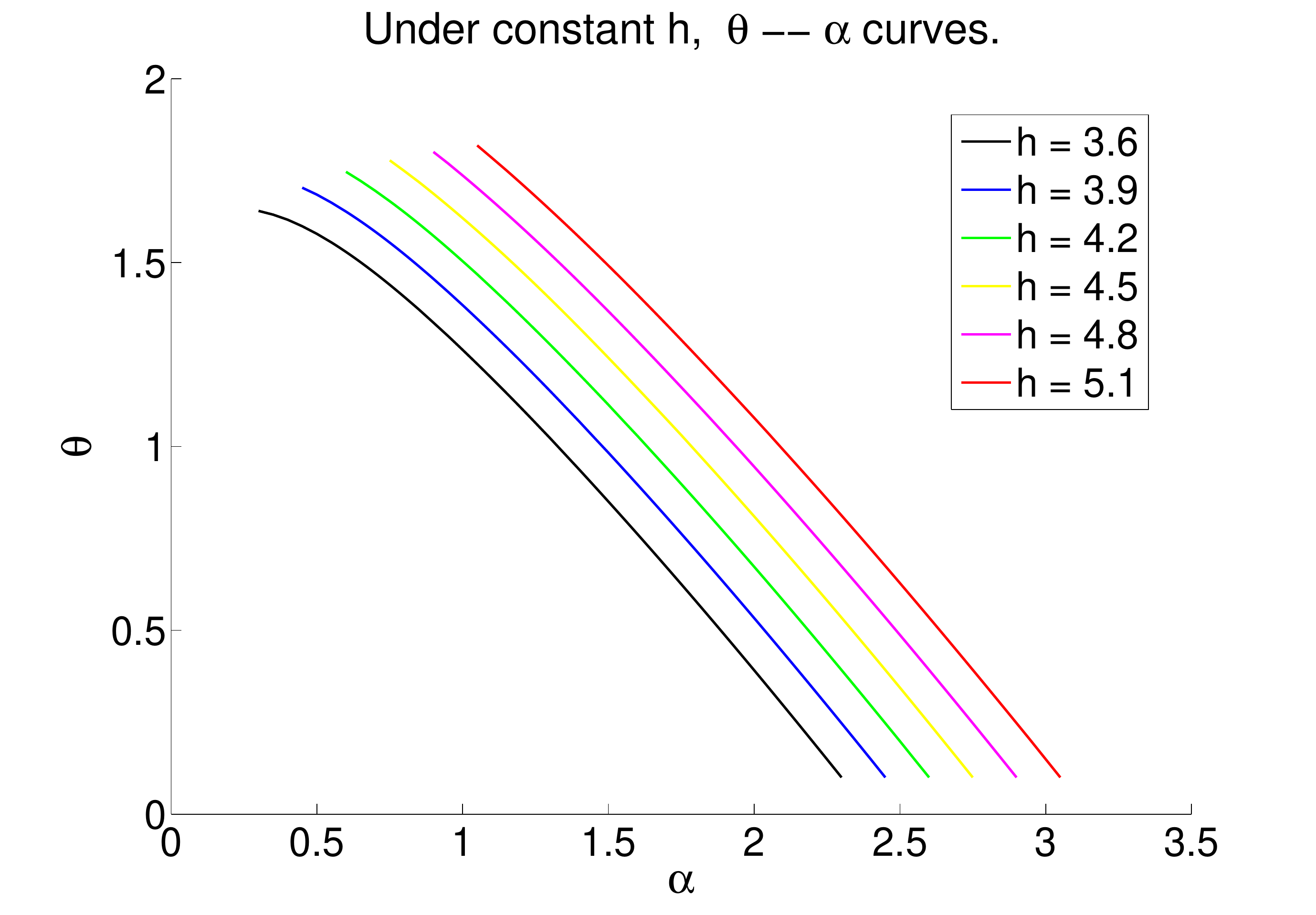}
\includegraphics[width=1.9in]{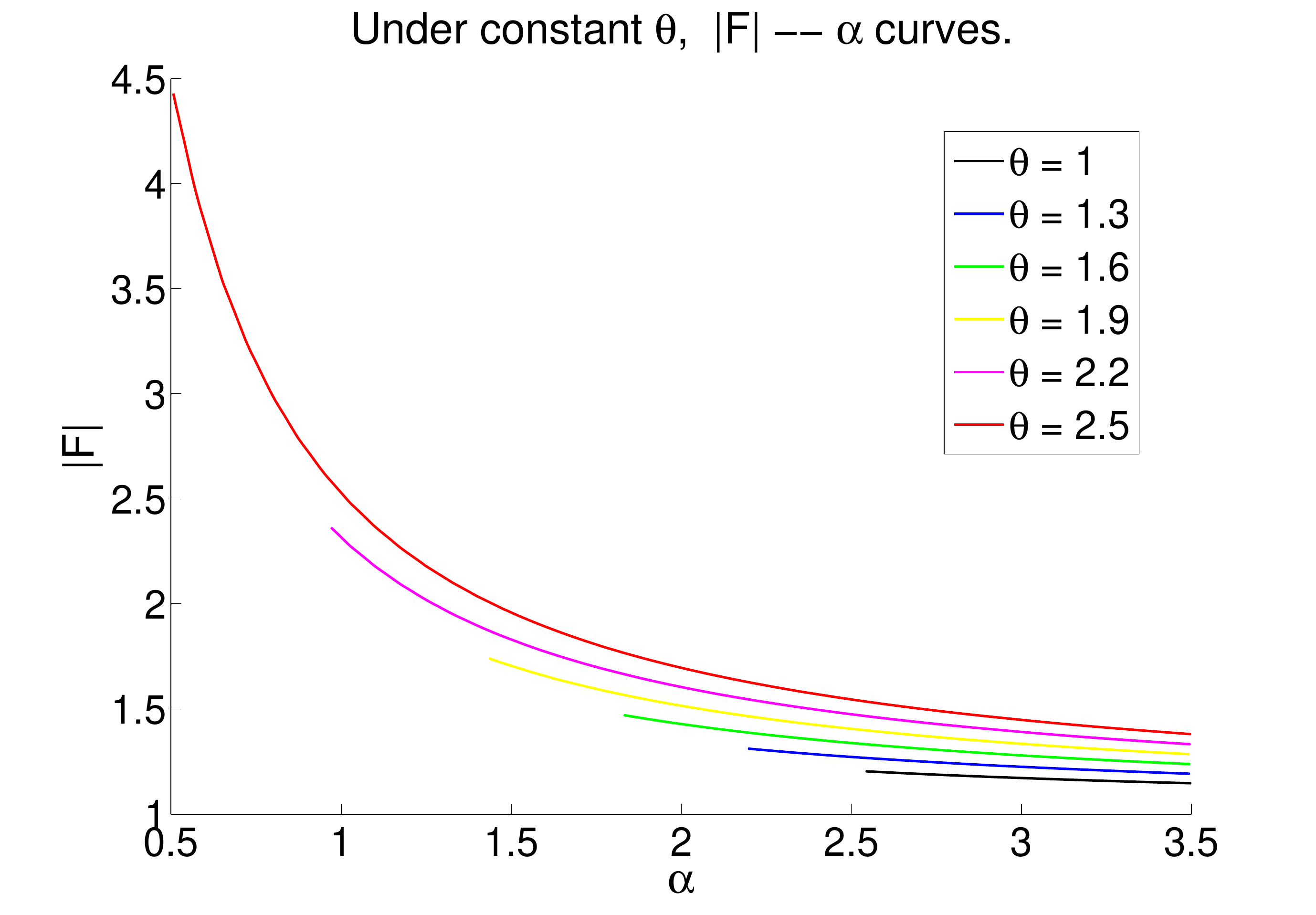}
\includegraphics[width=1.9in]{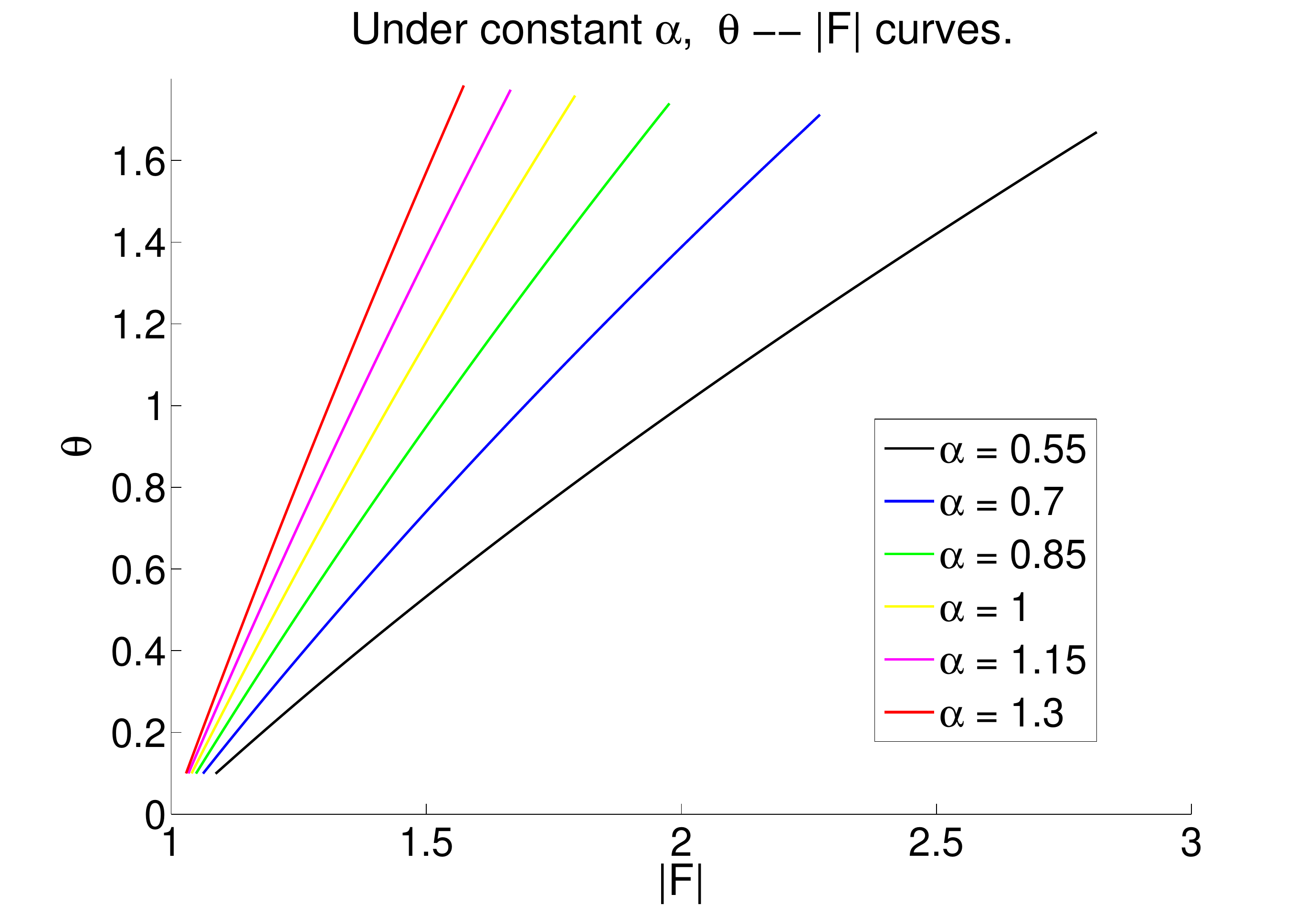}
\includegraphics[width=1.9in]{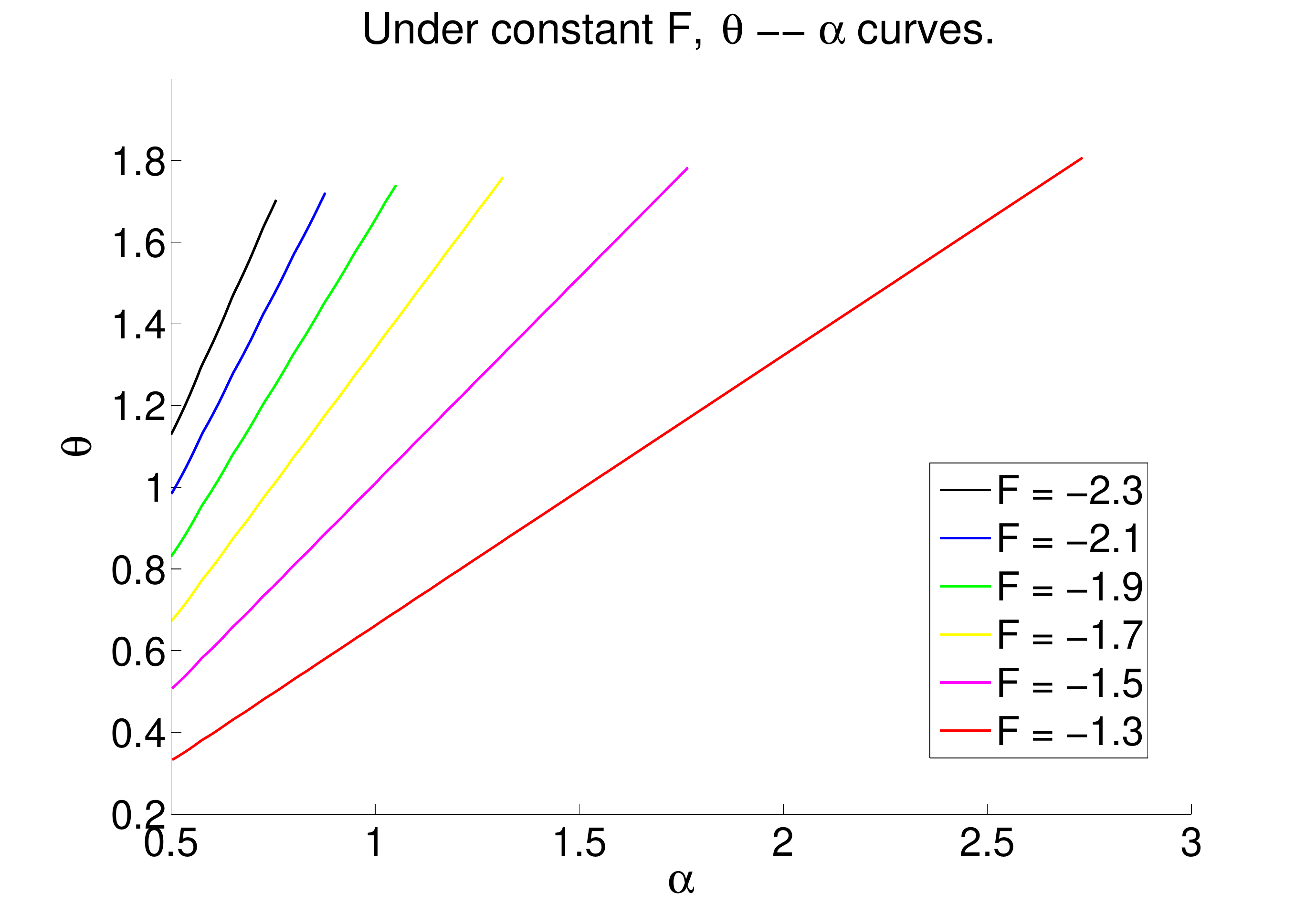}
\includegraphics[width=1.9in]{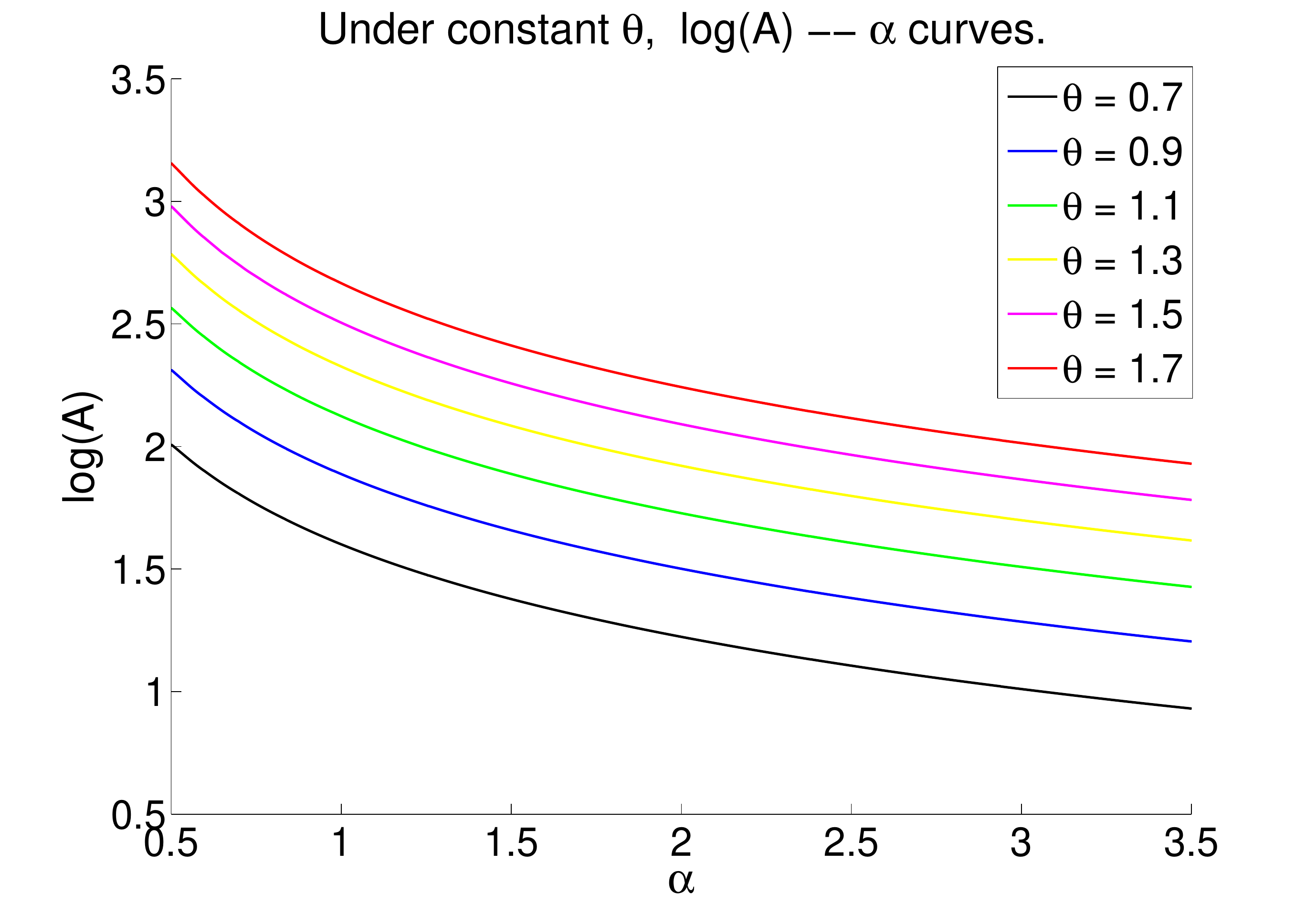}
\includegraphics[width=1.9in]{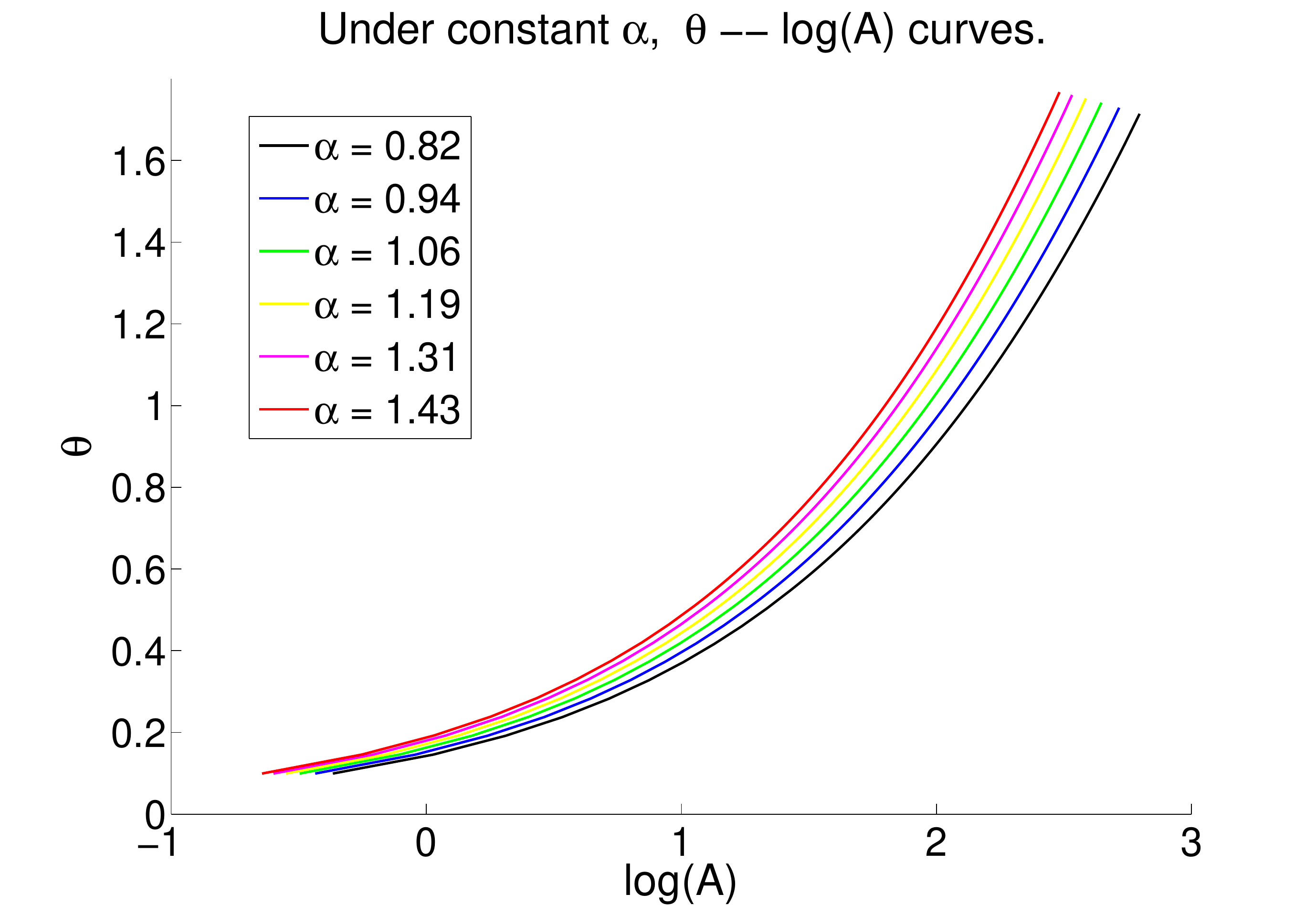}
\includegraphics[width=1.9in]{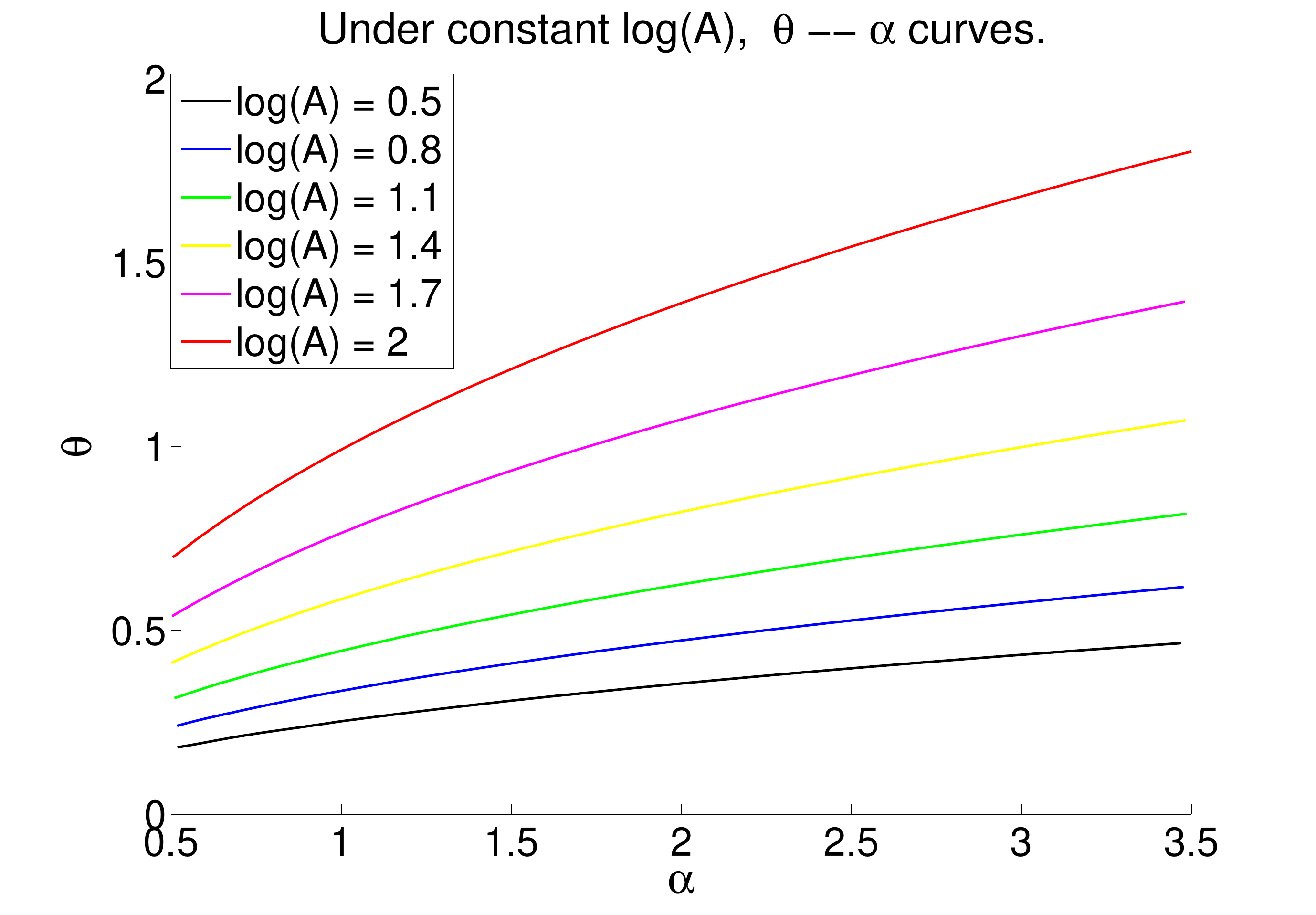}
\end{center}
\caption{Various functional relationships, e.g.,
``the equation of state'', among $\big(|F|,\alpha,h\big)$,
$\big(|F|,\theta,h\big)$, and $\big(\theta,\alpha,h\big)$ in the top row,
different views among $\big(|F|,\theta,\alpha\big)$ in the
second row,
and among $\big(\ln(\mathcal{A}),\theta,\alpha\big)$ in the third row.}
\label{fig_2}
\end{figure}

%

Fig. \ref{fig_2} shows various forms of ``the equation of state'',
e.g., relationships among
the triplets $\big(\alpha,|F_{\alpha}|=-F_{\alpha},h\big)$,
$\big(|F_{\alpha}|,\theta,h\big)$,
and $\big(\alpha,\theta,h\big)$ in the first row;
among the triplet $\big(\alpha,|F_{\alpha}|,\theta\big)$ in the second row;
and among the triplet $\big(\mathcal{A},\theta,\alpha\big)$ in the third row.
The second row shows that the relation among $\big(\alpha,|F_{\alpha}|,\theta\big)$ is just like that among $\big(V,P,T\big)$ in ideal gas model:
Mean ecological activeness $\theta$ increases nearly linearly with the ecological force $|F_{\alpha}|$ for constant $\alpha$, and with the proportion $\alpha$ of the predator's population range over the prey's, for constant $|F_{\alpha}|$;
Ecological force $|F_{\alpha}|$ and the proportion $\alpha$ of population ranges are inversely related under constant mean activeness $\theta$.
And when $\theta=0$, $\alpha(F_{\alpha}+1)=0$.
Other features can be observed by looking into the details of each column.

The first column of Fig. \ref{fig_2} demonstrates that as the proportion $\alpha$ of population ranges increases, the ecological force $|F_{\alpha}|$ is alleviated (for given $h$-energy or ecological activeness $\theta$).
This is due to the positive relationship between the ecological force $|F_{\alpha}|$ and the prey's population range (as shown in Eq. \ref{F_alpha}).
Since $\alpha$ is the proportion of the predator's population range over the prey's,
$|F_{\alpha}|$ and $\alpha$ would be inversely related when any resource, $h$-energy, or activity, $\theta$, remains constant.
This fact means that on an iso-$h$ or iso-$\theta$ curve, when the proportion $\alpha$ is large, relatively less $h$-energy change is needed to reduce it.
The first column also demonstrates an inverse relationship between $\alpha$ and the total population range $\ln\mathcal{A}$ for any given $\theta$, which reflects the fact that as the proportion of the predator's population range over the prey's increase, the total population range of the two species would actually decrease.

The second column of Fig. \ref{fig_2} demonstrates that: the ecological force $|F_{\alpha}|$ and the total population range $\ln\mathcal{A}$ increases with the mean activeness $\theta$ (with given $h$-energy or $\alpha$).
This observation means that it would also take more $h$-energy to change the proportion $\alpha$ of the predator's population range over the prey's, if mean ecological activeness rises, and that more population range would be explored with more ecological activeness $\theta$.

The third column about the relation between $\theta$ and $\alpha$ is interesting:
Under constant $h$-energy, as the proportion $\alpha$ of population ranges increases, the ecological activeness $\theta$ decreases, in accordance with the drop in the total population range $\ln\mathcal{A}$ as shown in Fig \ref{fig_1}.
But when the total population range $\ln\mathcal{A}$ or the ecological force $F_{\alpha}$ is to remain constant, ecological activeness $\theta$ actually increases with $\alpha$.
This means that under constant resource ($h$-energy), the proportion $\alpha$ of the predator's population range over the prey's restricts mean ecological activeness.
But if we fix the ecological force or total population range (supplying more $h$-energy), an increase in predator's population range over prey's can increase ecological activeness.

%
%
%
%
%

\section{Liouville description in phase space}
\label{sec3}

	Nonlinear dynamics described by Eq. (\ref{theeqn}) has a linear,
first-order partial differential equation (PDE) representation
\begin{equation}
   \frac{\partial u(x,y,t)}{\partial t} = -\frac{\partial}{\partial x}\Big(
                 f(x,y) u(x,y,t)\Big)-\frac{\partial}{\partial y}\Big(
                 g(x,y;\alpha) u(x,y,t)\Big).
\label{thepde}
\end{equation}
A solution to (\ref{thepde}) can be obtained via the method of
characteristics, exactly via (\ref{theeqn}).   Eq.
(\ref{thepde}) sometime is called the Liouville equation for
the ordinary differential equations (\ref{theeqn}).  It
also has an adjoint:
\begin{equation}
           \frac{\partial v(x,y,t)}{\partial t} = f(x,y)\frac{\partial v(x,y,t)}{\partial x}
                           +g(x,y;\alpha)\frac{\partial v(x,y,t)}{\partial y}.
\label{thepde2}
\end{equation}
Note that while the orthogonality in  Eq. (\ref{invmeasure})
indicates that $\rho\big(H(x,y)\big)$ is a stationary solution to
Eq. (\ref{thepde2}), it is not a stationary invariant density
to (\ref{thepde}).

This is due to the fact that vector field $(f,g)$ is not divergence
free, but rather as in (\ref{non-df}) the scalar factor $G(x,y)=xy$.
Then it is easy to verify that $G^{-1}(x,y)\rho(H(x,y))$ is a stationary
solution to (\ref{thepde}):
\begin{equation}
         \frac{\partial}{\partial x} \left(f(x,y)\ \frac{\rho(H(x,y))}{G(x,y)} \right)
       +  \frac{\partial}{\partial y} \left(g(x,y)\ \frac{\rho(H(x,y))}{G(x,y)} \right) = 0.
\label{non-ergodic}
\end{equation}

\subsection{Entropy dynamics in phase space}

	It is widely known that a volume-preserving, divergence-free
conservative dynamics has a conserved entropy
$S[u(x,t)] =-\int_{\mathbb{R}} u(x,t)\ln u(x,t)\rd x$ \cite{andrey85}.
For conservative system like (\ref{theeqn}) which contains the
scalar factor $G(x,y)$, the Shannon entropy should be replaced by
the relative entropy with respect to $G^{-1}(x,y)$ (see Appendix \ref{sec6.2} for detailed calculation):
\begin{eqnarray}
   	&& \frac{\rd}{\rd t}\int_{\mathbb{R}^2} u(x,y,t) \ln\left(
                 \frac{u(x,y,t)}{G^{-1}(x,y)}\right) \rd x\rd y
= 0.
\end{eqnarray}
Such systems are called {\em canonical conservative} with respect
to $G^{-1}(x,y)$ in \cite{qian_jmp}.  In classical statistical physics, the
term $\int_{\mathbb{R}}  u\ln\big(u/G^{-1}\big) \rd x$ is called
free energy \cite{WangJinEvolution}; in information theory, Kullback-Leibler divergence.

	We can in fact show a stronger result, with arbitrary differentiable
$\Psi(\cdot)$ and $\rho(\cdot)$ over an arbitrary domain $\mathfrak{D}$
(see Appendix \ref{sec6.2}):
\begin{eqnarray}
   	&& \frac{\rd}{\rd t}\int_{\mathfrak{D}} u(x,y,t) \Psi\left(
                 \frac{u(x,y,t)}{G^{-1}(x,y)\rho(H(x,y))}\right) \rd x\rd y
\nonumber\\[6pt]
	&=& \int_{\partial\mathfrak{D}}  \left\{ u(x,y,t)  \Psi\left(
                 \frac{u(x,y,t)}{G^{-1}(x,y)\rho(H)}\right) \big(f,g\big) \right\} \times (\rd x,\rd y).
\label{eq34}
\end{eqnarray}
Therefore, if $\mathfrak{D}=\mathfrak{D}_h$, then
$\partial\mathfrak{D}=\Gamma_{H=h}$, and the integral on the right-hand-side of
(\ref{eq34}) is always zero.  In other words, in
conservative dynamics like (\ref{theeqn}), it is the support
$\mathfrak{D}\subset \mathbb{R}^2$ on which $u(x,y,t)$
is observed that determines whether a system is
invariant; not the initial data $u(x,y,0)$ \cite{qian_pla}.

\subsection{Relation between $\mathcal{A}$, Shannon entropy, and
relative entropy}

	Since a ``state'' is defined as an entire orbit, it is natural to change the
coordinates from $(x,y)$ to
$(h,s)$ according to the solution curve to (\ref{theeqn}), where
we use $s$ to denote time, $0\le s\le \tau(h,\alpha)$.  We have
\begin{equation}
    \left(\frac{\partial x}{\partial t}\right)_{H=h} = x(1-y),  \  \
    \left(\frac{\partial y}{\partial t}\right)_{H=h} = \alpha y(x-1);
\end{equation}
\begin{equation}
     \left(\frac{\partial x}{\partial h}\right)_s \left(\alpha - \frac{\alpha}{x}\right)
       +  \left(\frac{\partial y}{\partial h}\right)_s
                 \left(1 - \frac{1}{y}\right)  = 1.
\end{equation}
Therefore:
\begin{equation}
    \det\left[\frac{D(x,y)}{D(h,s)}\right] =  xy = G(x,y).
\label{det}
\end{equation}
Then, the generalized relative entropy can be expressed as
\begin{eqnarray}
        && \int_{\mathfrak{D}_h} u(x,y,t) \Psi\left(
                 \frac{u(x,y,t)}{G^{-1}(x,y)\rho(H(x,y))}\right) \rd x\rd y
\nonumber\\
	&=& \int_{h_{min}}^{h} \rd \eta \int_0^{\tau(\eta,\alpha)}
               \frac{u\big(x(s),y(s),t\big)}{G^{-1}\big(x(s),y(s)\big)} \Psi\left(
                 \frac{u\big(x(s),y(s),t\big)}{G^{-1}\big(x(s),y(s)\big)
               \rho(\eta)}\right) \rd s
\nonumber\\
	&=& \int_{h_{min}}^h  \rho(\eta) \rd \eta \int_0^{\tau(\eta,\alpha)}
               \wtu\big(x(s),y(s),t;\eta\big)
               \Psi \Big( \wtu\big(x(s),y(s),t;\eta\big)\Big) \rd s
\nonumber\\
	&=&   \int_{h_{min}}^h  \Omega_B(\eta) \rho(\eta) \rd\eta,
\label{eq39}
\end{eqnarray}
in which
\[
            \wtu(x,y,t;h) =  \frac{u\big(x,y,t\big)}{G^{-1}\big(x,y\big)
               \rho(h)}.
\]
and
\begin{equation}
          \Omega_B(h) =  \oint_{\Gamma_{H=h}}
               \wtu\big(x,y,0;h\big)
               \Psi \Big( \wtu\big(x,y,0;h\big)\Big) \rd\ell.
\end{equation}
 $\Omega_B(h)$ is known as Boltzmann's entropy in classical
statistical mechanics.  We see that the $\mathcal{A}$ introduced
in Sec. \ref{sec2} is the simplest case of the generalized
relative entropy in (\ref{eq34}) with $\rho=\Psi=1$, and
$u(x,y,0)=G^{-1}(x,y)$.  Then
$\Omega_B(h)=\rd \mathcal{A}_{\mathfrak{D}_h}/\rd h$.
Gibbs' canonical ensemble chooses $\rho(h)=e^{-h/\theta}$.

The dynamics (\ref{theeqn}) is not ergodic in the $xy$-plane;
it does not have a unique invariant measure, as indicated by the
arbitrary $\rho(H)$ in Eq. (\ref{non-ergodic}).  However, the function
$G(x,y)$, as indicated in Eqs.  (\ref{t2that}) and (\ref{det}),
is the unique invariant measure on each ergodic invariant
submanifold $\Gamma_{H=h}$.  It is
non-uniform with respect to Lebesgue measure.  On the
ergodic invariant manifold $\Gamma_{H=h}$:
$G(x,y) \rd t \leftrightarrow \rd\ell$. To see the difference between
the Lebesgue-based average and invariant-measure based average,
consider a simple
time-varying exponentially growing population: $\frac{\rd u(t)}{\rd t} =
r(t) u(t)$.  The regular  time average of the per capita growth rate is
\[
       \frac{1}{\tau}\int_0^{\tau}  \frac{1}{u(t)}\frac{\rd u}{\rd t} \rd t
         = \frac{1}{\tau}\ln\left(\frac{u(\tau)}{u(0)}\right)
         = \frac{1}{\tau}\int_0^{\tau} r(t)dt.
\]
The Lebesgue-based average is an ``average growth rate per
average capita''
\[
   \frac{\displaystyle
            \int_0^{\tau} \frac{1}{u(t)}\frac{\rd u}{\rd t}\ u(t)\rd t }
                 {\displaystyle   \int_0^{\tau} u(t)\rd t }
         = \frac{ \displaystyle    \int_0^{\tau} r(t) u(t) \rd t }
                  { \displaystyle \int_0^{\tau}  u(t) \rd t }.
\]
In cyclic population dynamics, this latter quantity corresponds to the $G\big(x(t),y(t)\big)$ weighted per capita growth rate or ``kinetic energy"
\begin{eqnarray}
&&\frac{\displaystyle\int_0^{\tau(h,\alpha)} \left(\frac{\rd \ln(y)}{\rd t}\right)
G\big(x(t),y(t)\big) \rd t} {\displaystyle\int_0^{\tau(h,\alpha)}
G\big(x(t),y(t)\big) \rd t}
=\frac{\displaystyle\int_0^{\tau(h,\alpha)} - \left(\frac{\rd \ln(x)}{\rd t}\right)
 G\big(x(t),y(t)\big)  \rd t} {\displaystyle\int_0^{\tau(h,\alpha)}  G\big(x(t),y(t)\big) \rd t}
\nonumber\\[9pt]
=&&\frac{\displaystyle\int_0^{\tau(h,\alpha)} x\left(\frac{\partial H}{\partial x}\right) G\big(x(t),y(t)\big)  \rd t} {\displaystyle\int_0^{\tau(h,\alpha)}  G\big(x(t),y(t)\big) \rd t}
=\frac{\displaystyle\int_0^{\tau(h,\alpha)} y \left(\frac{\partial H}{\partial y} \right)
G\big(x(t),y(t)\big) \rd t} {\displaystyle\int_0^{\tau(h,\alpha)}
G\big(x(t),y(t)\big) \rd t}.
\end{eqnarray}

\section{Stochastic description of finite populations}
\label{sec4}

	In this section, we show that the conservative dynamics
in (\ref{theeqn}) is an emergent caricature of a robust stochastic
population dynamics.  This material can be found in many texts,
e.g., \cite{kurtz}. But for completeness, we shall give a brief
summary.

Assume the populations of the prey and the predator, $M(t)$ and $N(t)$,
reside in a spatial region of size $\Omega$.
The discrete stochastic population dynamics follows a two-dimensional,
continuous time birth-death process with transition
probability rate
\begin{eqnarray}
     && \Pr\Big\{ M(t+\Delta t) =k, N(t+\Delta t) =\ell \
                \Big|M(t) =m, N(t)=n\Big\}
\nonumber\\[6pt]
                 &=&  \left(  m\delta_{k,m+1}
                 +\dfrac{1}{\Omega} mn \delta_{k,m-1}
                 +\dfrac{\alpha}{\Omega} nm \delta_{\ell,n+1}
                 +\alpha n\delta_{\ell,n-1} \right)\Delta t + o(\Delta t).
\label{bdp}
\end{eqnarray}
The discrete stochastic dynamics has an invariant measure:
\begin{eqnarray}
     {\rm Pr} ^{ss}\Big\{ M = m, N = n \Big\}
     =  \dfrac{1}{mn}.
\label{bdp}
\end{eqnarray}
Then
\begin{eqnarray}
     p_{m,n}(t+\Delta t) &=& p_{m,n}(t)\left[1-\left(m+\dfrac{1}{\Omega}mn+\dfrac{\alpha}{\Omega} nm + \alpha n\right)\Delta t\right]
\nonumber\\[6pt]
     &+& p_{m-1,n}(t)\Big[(m-1)\Delta t\Big] + p_{m+1,n}(t)
            \left[\dfrac{1}{\Omega}(m+1)n\Delta t\right]
\nonumber\\[6pt]
     &+& p_{m,n-1}(t)\left[\dfrac{\alpha}{\Omega}(n-1)m\Delta t\right]
               + p_{m,n+1}(t)\Big[\alpha(n+1)\Delta t\Big].
\nonumber
\end{eqnarray}
That is
\begin{eqnarray}
     \dfrac{p_{m,n}(t+\Delta t)-p_{m,n}(t)}{\Delta t} &=& -m \Big[p_{m,n}(t)-p_{m-1,n}(t) \Big] -p_{m-1,n}(t)
\nonumber\\[6pt]
     &+& \dfrac{1}{\Omega} mn \Big[p_{m+1,n}(t)-p_{m,n}(t) \Big] + \dfrac{1}{\Omega} n p_{m+1,n}(t)
\nonumber\\[6pt]
     &-& \dfrac{\alpha}{\Omega} nm  \Big[p_{m,n}(t)-p_{m,n-1}(t) \Big] - \dfrac{\alpha}{\Omega} m p_{m,n-1}(t)
\nonumber\\[6pt]
     &+& \alpha n \Big[p_{m,n+1}(t)-p_{m,n}(t) \Big] + \alpha p_{m,n+1}(t).
\label{eq49}
\end{eqnarray}

For a very large $\Omega$, the population {\em densities} at
time $t$ can be approximated by continuous random variables as $X(t)=\Omega^{-1}M(t)$ and $Y(t)=\Omega^{-1}N(t)$.   Then
Eq. (\ref{eq49}) becomes a partial differential equation by
setting $x=m/\Omega$, $y=n/\Omega$, and $u(x,y,t)=p_{m,n}(t)/\Omega$:
\begin{eqnarray}
     \dfrac{\partial u}{\partial t} &=& -x \dfrac{\partial u}{\partial x} + \dfrac12 \Omega^{-1}x \dfrac{\partial^2 u}{\partial x^2}
     - u + \Omega^{-1}\dfrac{\partial u}{\partial x}
\nonumber\\[6pt]
     &+& xy\dfrac{\partial u}{\partial x} + \dfrac12 \Omega^{-1} xy \dfrac{\partial^2 u}{\partial x^2}
     + y u + \Omega^{-1} y \dfrac{\partial u}{\partial x}
\nonumber\\[6pt]
     &-& \alpha xy \dfrac{\partial u}{\partial y} + \dfrac\alpha2  \Omega^{-1} xy \dfrac{\partial^2 u}{\partial y^2}
     - \alpha x u + \alpha \Omega^{-1} x \dfrac{\partial u}{\partial y}
\nonumber\\[6pt]
     &+& \alpha y \dfrac{\partial u}{\partial y} + \dfrac\alpha2  \Omega^{-1}y \dfrac{\partial^2 u}{\partial y^2}
     + \alpha u + \alpha \Omega^{-1}\dfrac{\partial u}{\partial y} + o(\Omega^{-1}).
\nonumber
\end{eqnarray}
Rearranging the terms and writing $\epsilon=\Omega^{-1}$, we can perform the Kramers-Moyal expansion to obtain:
\begin{eqnarray}
           \frac{\partial u(x,y,t)}{\partial t} &=& \nabla\cdot
           \Big( \epsilon\mD(x,y) \nabla u -  \mF(x,y) u\Big) + \dfrac{\epsilon}{2}\left((y+1) \dfrac{\partial u}{\partial x} + \alpha(x+1) \dfrac{\partial u}{\partial y}\right) + o(\epsilon)
\nonumber\\
           &=& \epsilon \sum_{\xi=x,y}\sum_{\zeta=x,y}
             \frac{\partial^2}{\partial\xi\partial\zeta} D_{\xi\zeta}(x,y)u(x,y,t)
            - \nabla\cdot \Big(\mF(x,y)\ u\Big),
\label{thefpe}
\end{eqnarray}
with drift $\mF(x,y)=\big(f(x,y),g(x,y;\alpha)\big)^T$
and symmetric diffusion matrix
\begin{equation}
   \mD(x,y) = \left(\begin{array}{cc}
				D_{xx}(x,y) &  D_{xy}(x,y)   \\
			      D_{yx}(x,y)	&  D_{yy}(x,y)
                     \end{array} \right)
                   = \frac{1}{2}\left(\begin{array}{cc}
				x(1+ y) &    0   \\
						0	& \alpha y (x+1)
                     \end{array} \right).
                     \nonumber
\end{equation}

	Eq. (\ref{thefpe}) should be interpreted as a
Fokker-Plank equation for the probability density function
$u(x,y,t)\rd x\rd y = \Pr\big\{x<X(t)\le x+\rd x, y<Y(t)\le y+\rd y \big\}$.
It represents a continuous stochastic process $\big(X(t),Y(t)\big)$
following It\={o} integral \cite{linda_allen,grasman,kurtz}:
\begin{eqnarray}
	\rd X(t) &=& X(1-Y)\rd t + \epsilon^{\frac{1}{2}} \sqrt{X(1+Y)}\ \rd W_1(t)
\nonumber\\[-7pt]
\label{sde}
\\[-7pt]
 	\rd Y(t) &=& \alpha Y(X-1) \rd t + \epsilon^{\frac{1}{2}}\sqrt{\alpha Y(X+1)}\ \rd W_2(t)
\nonumber
\end{eqnarray}
It is important to recognize that in the limit of $\epsilon\rightarrow 0$,
the dynamics described by Eq. (\ref{thefpe}) is reduced to
that in Eq. (\ref{thepde}), which is equivalent to Eq. (\ref{theeqn})
via the method of characteristics.


\subsection{Potential-current decomposition}

It can be verified that the stationary solution to (\ref{thefpe}) is actually
$G^{-1}(x,y)=(xy)^{-1}$, which is consistent with the discrete case (cf. Eq. \ref{bdp}), and also a stationary solution to the Liouville equation Eq. (\ref{thepde}).

As suggested in \cite{wangjin,qian_pla}, the
right-hand-side of  Eq. (\ref{thefpe}) has a natural decomposition:
\begin{eqnarray}
  && \nabla\cdot
           \Big( \epsilon\mD(x,y) \nabla u -  \mF(x,y) u\Big) + \dfrac{\epsilon}{2}\left((y+1) \dfrac{\partial u}{\partial x} + \alpha(x+1) \dfrac{\partial u}{\partial y}\right)
\nonumber\\
   &=& \nabla\cdot
           \Big[ \epsilon\mD(x,y) \nabla u -  \Big(\mF(x,y) - \epsilon\mD(x,y) \nabla \ln G(x,y) \Big) u  \Big]
\nonumber\\
   &= & \nabla\cdot
           \Big[ \epsilon\mD \Big(u\nabla\ln u +u \nabla\ln G \Big) - \mF u  \Big]
\nonumber\\
   &=& \epsilon\nabla\cdot \mD
          u \nabla \Big(\ln \big( G \ u \big) \Big)
              - \nabla\cdot \big( \mF \ u \big)
\end{eqnarray}
in which the first term is a self-adjoint differential operator and
the second is skew-symmetric \cite{qian_jmp}.  The equation
from the first line to the second uses the fact
$\nabla\ln G = -\big(x^{-1},y^{-1}\big)$, thus
$\mD\nabla\ln G = -\frac{1}{2}\big( (y+1),\alpha(x+1)\big)$.
In terms of the stochastic differential equation in divergence form, this
decomposition corresponds to:
\begin{equation}
       \left(\begin{array}{c}
                  \rd X \\[5pt] \rd Y \end{array}\right) = -\epsilon\mD
                      \nabla \ln G
                + G \left(\begin{array}{c}
                             -H_y \\[5pt]  H_x \end{array}\right)
               + \epsilon^{\frac{1}{2}} \sqrt{2\mD} \left(\begin{array}{c}
                       \rd W_1(t) \\[5pt]   \rd W_2(t) \end{array}\right).
\label{newsde}
\end{equation}
Under this non-It\={o} interpretation of the stochastic
differential equation, the finite population with fluctuations
(i.e., $\epsilon\neq0$) is unstable when $x, y>0$.
The system behaves as an unstable focus as shown in Fig. \ref{fig_3}.
The eigenvalues at the fixed point $\big(1+\epsilon,1-\epsilon\big)$
are $\pm i\sqrt{\alpha}+\frac{1}{2}\epsilon(\alpha+1)$,
corresponding to the unstable nature of the stochastic system.

\begin{figure}[tbh]
\begin{center}
\includegraphics[width=2.5in]{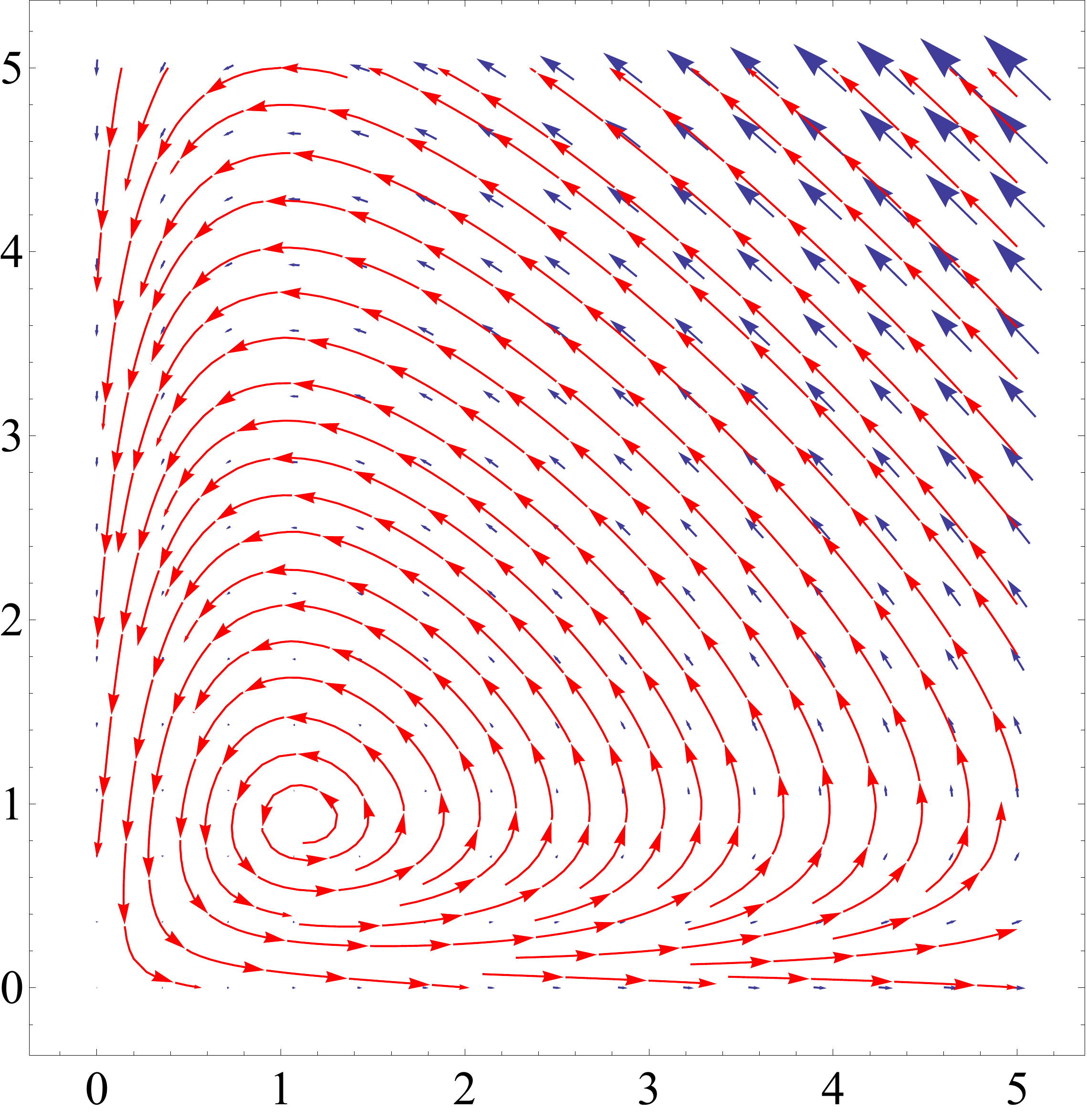}
\end{center}
\caption{With fluctuation ($\epsilon=0.1$), the deterministic part
of system (\ref{newsde}).  Red curves denote phase flow and blue arrows denote the strength of the vector field.}
\label{fig_3}
\end{figure}

On the other hand, the potential-current decomposition reveals that the
system (\ref{theeqn}) will be structurally stable in terms of the
stochastic model:  Any perturbation of the model system will yield
corresponding conserved dynamics close to (\ref{theeqn}).
The conservative ecology is a robust emergent phenomenon.

Equations such as (\ref{sde}) and (\ref{newsde}) are
not mathematically well-defined until an precise meaning
of integration
\begin{equation}
         X(t) =  \int_0^t    b\big(W(t)\big)  \rd W(t)
\label{wt}
\end{equation}
is prescribed.  This yields different stochastic processes
$X(t)$ whose corresponding probability density function
$f_{X(t)}(x,t)$ follow different  linear partial differential
equations.  The fundamental solution to any
partial differential equation (PDE), however, provides a
Markov transition probability; there is no ambiguity
at the PDE level.  On the other hand, the only interpretation
of (\ref{wt}) that provides a Markovian stochastic process that is
non-anticipating is that of It\={o}'s \cite{Gardiner}.  The differences in
the interpretations of (\ref{wt}) become significant only in the
modeling context, when one's intuition expects that
$E\big[X(t)\big]=0$ even for interpretations other then It\={o}'s.

\subsection{The slowly fluctuating \boldmath{$H_t=H\big(X(t),Y(t)\big)$}}

	With the $\big(X(t),Y(t)\big)$ defined in (\ref{thefpe}) and
(\ref{sde}), let us now consider the stochastic functional
\begin{eqnarray}
	\rd H\big(X(t),Y(t)\big) &=& \alpha\left(1-\frac{1}{X}\right)\rd X
                + \left(1-\frac{1}{Y}\right) \rd Y
              + \frac{1}{2}
	\left(\frac{(\rd X)^2}{X^2}  + \frac{(\rd Y)^2}{Y^2}  \right)
\nonumber\\
	&=& \alpha \epsilon^{\frac{1}{2}} \left(
                    \frac{(X-1)^2(1+Y)}{X}
                + \frac{(Y-1)^2(X+1)}{Y}\right)^{\frac{1}{2}}
                            \rd W(t)
\nonumber\\
	&& + \frac{\epsilon}{2}\left( \frac{(1+Y)}{X}
                      +\frac{\alpha (X+1)}{Y}\right)\rd t
\end{eqnarray}
Therefore, for very large populations, i.e., small $\epsilon$, this suggests a
separation of time scales between the cyclic motion
on $\Gamma_{H=h}$ and slow, stochastic level crossing
$H_t$.  The method of averaging is applicable here
\cite{fw_book,zhuwq}:
\begin{equation}
        \rd H_t =  \epsilon b(H_t)\rd t + \epsilon^{\frac{1}{2}}  A(H_t)\rd W(t),
        \label{diffh}
\end{equation}
with
\begin{eqnarray}
   b(h) &=&  \frac{1}{2} \left\langle \frac{(1+y)}{x}
                      +\frac{\alpha (x+1)}{y}\right\rangle^{\Gamma_{H=h}},
\\
	A(h) &=&  \alpha \left\langle \left(
                    \frac{(x-1)^2(1+y)}{x}
                + \frac{(y-1)^2(x+1)}{y}\right)^{\frac{1}{2}}  \right\rangle^{\Gamma_{H=h}},
\end{eqnarray}
where $\left\langle \psi(x,y) \right\rangle^{\Gamma_{H=h}} = \left\langle \psi(x,y) \right\rangle^t$
denotes the average of $\psi(x,y)$ on the level set $\Gamma_{H=h}$.
Then, using the It\={o} integral, the distribution of $H_t$ follows a Fokker-Planck equation:
\begin{equation}
    \dfrac{\partial p(H,t)}{\partial t}
    = - \epsilon \dfrac{\partial}{\partial H}\big(b(H) p\big)
    + \dfrac{\epsilon}{2} \dfrac{\partial^2}{\partial H} \Big(A^2(H) p\Big).
    \label{varyingH}
\end{equation}
And the stationary solution for Eq. (\ref{varyingH}) is:
\begin{eqnarray}
p^{ss}(H) = \dfrac{1}{A^2(H)} \exp\left( {\displaystyle 2\int_{H_0}^H \dfrac{b(h)}{A^2(h)} \rd h} \right).
\end{eqnarray}
The steady state distributions of $H$ under different $\alpha$'s are
shown in Fig. \ref{fig_3}.
The steady state distribution $p^{ss}(H)$ does not depend on the volume size $\Omega=\epsilon^{-1}$.

\begin{figure}[tbh]
\begin{center}
\includegraphics[width=3.5in]{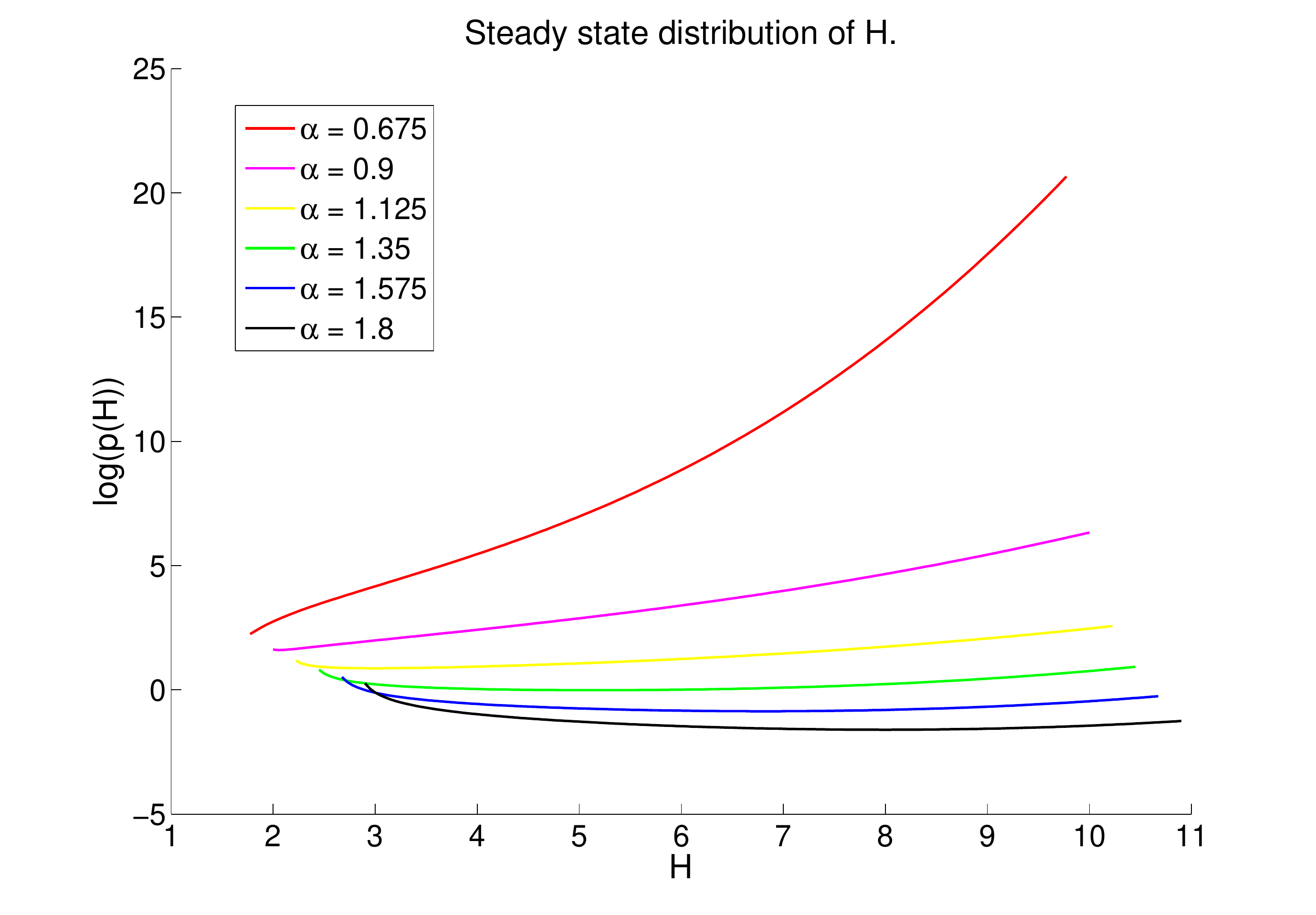}
\end{center}
\caption{Under different values of $\alpha$, the steady state distribution $p^{ss}(H)$ with respect to $H$ in logarithmic scale.
The slowly fluctuating ``energy" $H$ ranges from $\alpha+1$ to infinity.
Its steady state distribution $p^{ss}(H)$ eventually increases without bound as $H$ increases.}
\label{fig_3}
\end{figure}

When $H$ is big enough,
$p^{ss}(H)$ increases with $H$ without bound, since $b(H)$ is a positive increasing function.
Hence, $p^{ss}(H)$ is not normalizable on the entire $\mathbb{R}$,
reflecting the unstable nature of the system.
The fluctuation $A(H)$ approaches zero when $H$ approaches $\alpha+1$.
Consequently, the absorbing effect at $H=\alpha+1$ makes $p^{ss}(\alpha+1)$ another possible local maximum.


\section{Discussion}
\label{sec5}

	It is usually an obligatory step in understanding an ODE
$\dot{x}=f(x;\alpha,\beta)$ to analyze the dependence of a stea\rd y
state $x^*$ as an implicit function  of the parameters $(\alpha,\beta)$
\cite{jdmurray}.  One of the important phenomena in this regard is
the Thom-Zeeman catastrophe \cite{jdmurray,aqtw}.  From this
broad perspective, the analysis developed by  Helmholtz and
Boltzmann in 1884 is an analysis of the geometry of a ``non-constant
but stea\rd y solution'', as a function of its
parameter(s) and initial conditions.   In the context of LV
equation (\ref{theeqn}),  the geometry is characterized by
the area encircled by a periodic solution, $\Gamma_{H=h}$,
where $h$ is specified by the the initial data:
$\mathcal{A}\big(\mathfrak{D}_h\big)=\mathcal{A}\big(h,\alpha
\big)$.  The celebrated Helmholtz theorem \cite{gallavotti,mcampisi_05} then
becomes our Eq. (\ref{dheq28})
\begin{equation}
        \rd h  =  \frac{\displaystyle \rd\mathcal{A}
                  - \left(\frac{\partial\mathcal{A}}{\partial\alpha}
                    \right)_{h} \rd\alpha } {\displaystyle
                       \left(\frac{\partial\mathcal{A}}{\partial h}
                    \right)_{\alpha} }
                    =  \theta(h,\alpha)\
                \rd\ln\mathcal{A}
                  - F_{\alpha}(h,\alpha) \rd\alpha.
\label{dheq}
\end{equation}
Since Eq. (\ref{theeqn}) has a conserved
quantity $H$, Eq. (\ref{dheq}) can, and should be, interpreted as an extended
$H$ conservation law, beyond the dynamics along a single
trajectory, that includes both variations in $\alpha$ and in $h$.
The partial derivatives in (\ref{dheq}) can be shown
as time averages of ecological activeness
$\left\langle \ln x (\partial H/\partial\ln x) \right\rangle^{t}$
or $\left\langle \ln y (\partial H/\partial\ln y) \right\rangle^{t}$,
and variation in the prey's population $\left\langle x - \ln x \right\rangle^{t}$.
Those conjugate variables, along with parameter $\alpha$,
conserved quantity $H$, 
and encompassed area $\ln \mathcal{A}$
constitutes a set of ``state variables" describing the state of an
ecological system in its stationary, cyclic state.
This is one of the essences of Boltzmann's statistical mechanics \cite{gallavotti}.

For the monocyclic Lotka-Volterra system, the dynamics are relatively simple.
Hence, the state variables have monotonic relationships,
the same as that observed in ideal gas models.
When the system's dynamics become more complex (e.g. have more than one attractor, Hopf bifurcation),
relations among the state variables will reflect that complexity
(e.g. develop a cusp, exhibiting a phase transition in accordance
\cite{aqtw}).

When the populations of predator and prey are finite,
the stochastic predator-prey dynamics is unstable.
This fact is reflected in the non-normalizable steady state distribution
$G^{-1}(x,y)$ on $\mathbb{R}^{2+}$, and the
destabilizing effect of the gradient dynamics
in the potential-current decomposition.  This is particular
to the LV model we use; it is not a problem for the general theory if we study a more realistic model as in \cite{WangJinEcology}.
Despite the unstable dynamics, the stochastic model system is structurally stable: its dynamics persists under sufficiently small perturbations.
This implies conservative dynamical systems like (\ref{theeqn})
are meaningful mathematical models, when interpreted correctly,
for ecological realities.


	Indeed, all ecological population dynamics can be represented by
birth-death stochastic processes \cite{kurtz}.
Except for systems with detailed balance, which rarely holds true,
almost all such dynamics have underlying cyclic, stationary
conservative dynamics.  The present work shows that a hidden
conservative ecological dynamics can be revealed through
mathematical analyses.   To
recognize such a conservative ecology, however, several novel
quantities need to be defined, developed, and becoming
a part of ecological vocabulary.
This is the intellectual legacy of Helmholtz's and Boltzmann's
mechanical theory of heat \cite{gallavotti_2}.

\vskip 0.3cm

	{\bf Authors' contributions.}  Y.-A. M. and H. Q. contributed equally to this work.

	{\bf Competing interests.}  We declare there are no competing interests.

	{\bf Acknowledgements.}  We thank R.E. O'Malley, Jr.
and L.F. Thompson for carefully reading the manuscripts and
many useful comments.

\appendix

\section{Population temporal variations}
\label{app-A}
\label{sec:AA}
\setcounter{equation}{0}
\renewcommand\theequation{A\arabic{equation}}

\begin{eqnarray}
       \int_0^{\tau} \big(x(t)-1\big)^2 \rd t &=& \frac{1}{\alpha}
        \int_0^{\tau} \left(\frac{x-1}{y}\right)\rd y(t)
\nonumber\\
	&=&    \frac{1}{\alpha} \int_0^{\tau} (x-1)\rd\Big(
                       \alpha x(t)+y(t)-\alpha\ln x(t)\Big)
\nonumber\\
	&=&   \frac{1}{\alpha} \int_0^{\tau} (x-1)\rd y(t)
\nonumber\\
	&=&   \frac{1}{\alpha} \int_0^{\tau} x(t)\ \rd y(t) \ = \
            \frac{\hat{\mathcal{A}}}{\alpha}.
\end{eqnarray}
Similarly
\begin{equation}
    \int_0^{\tau} \big(y(t)-1\big)^2 \rd t = -\alpha \int_0^{\tau} y(t)\ \rd x(t)
             \ = \ \alpha\hat{\mathcal{A}}.
\end{equation}

\section{Dynamics of relative entropy and generalized relative entropy}
\label{app-B}
\label{sec6.2}
\setcounter{equation}{0}
\renewcommand\theequation{B\arabic{equation}}

Using the divergence theorem and noting that $\nabla\cdot\Big(G^{-1}f,G^{-1}g\Big)=0$, we obtain for the time evolution of the relative entropy:
\begin{eqnarray}
   	&& \frac{\rd}{\rd t}\int_{\mathbb{R}^2} u(x,y,t) \ln\left(
                 \frac{u(x,y,t)}{G^{-1}(x,y)}\right) \rd x\rd y
\nonumber\\[6pt]
     &=&
        \int_{\mathbb{R}^2}\frac{\partial u(x,y,t)}{\partial t} \left[\ln\left(
                 \frac{u(x,y,t)}{G^{-1}(x,y)}\right) + 1 \right] \rd x\rd y
\nonumber\\[6pt]
	&=& -\int_{\mathbb{R}^2} \nabla\cdot\Big( (f,g)u(x,y,t)\Big) \ln\left(
                 \frac{u(x,y,t)}{G^{-1}(x,y)}\right) \rd x\rd y
        +\frac{\partial}{\partial t}\int_{\mathbb{R}^2} u(x,y,t) \rd x\rd y
\nonumber\\[6pt]
	&=& -\int_{\mathbb{R}^2} \nabla\cdot\Big( (f,g)u(x,y,t)\Big) \ln\left(
                 \frac{u(x,y,t)}{G^{-1}(x,y)}\right) \rd x\rd y\
\nonumber\\[6pt]
	&=& -\int_{\mathbb{R}^2} \nabla\cdot\Big(G^{-1}f,G^{-1}g\Big) \times \left(\frac{u}{G^{-1}}\right)
                 \ln\left(\frac{u}{G^{-1}}\right) \rd x\rd y
\nonumber\\[6pt]
    &&    -\int_{\mathbb{R}^2} \Big(G^{-1}f,G^{-1}g\Big) \cdot \nabla\left[\left(\frac{u}{G^{-1}}\right)
                 \ln\left(\frac{u}{G^{-1}}\right) - \left(\frac{u}{G^{-1}}\right)\right] \rd x\rd y
\nonumber\\[6pt]
	&=& \int_{\mathbb{R}^2} \left[ \left(\frac{u}{G^{-1}}\right)
                          \ln\left(\frac{u}{G^{-1}}\right)-\left(\frac{u}{G^{-1}}\right)
                        \right]
           \nabla\cdot \Big(G^{-1}f,G^{-1}g\Big) \ \rd x\rd y
	 \  = \  0.
\end{eqnarray}
A more general result can be obtained in parallel for arbitrary
differentiable $\Psi(z)$ and $\rho(z)$:
\begin{eqnarray}
   	&& \frac{\rd}{\rd t}\int_{\mathfrak{D}} u(x,y,t) \Psi\left(
                 \frac{u(x,y,t)}{G^{-1}(x,y)\rho(H(x,y))}\right) \rd x\rd y
\nonumber\\[6pt]
	&=&  \int_{\mathfrak{D}} \frac{\partial u(x,y,t)}{\partial t} \left[
                         \Psi\left(
                 \frac{u}{G^{-1}\rho(H)}\right)  +
		    \frac{u}{G^{-1}\rho(H)} \Psi'\left(
                 \frac{u}{G^{-1}\rho(H)}\right) \right] \rd x\rd y
\nonumber\\[6pt]
	&=& -\int_{\mathfrak{D}} \nabla\cdot\Big( (f,g) u\Big) \left[
                         \Psi\left(
                 \frac{u}{G^{-1}\rho(H)}\right)  +
		    \frac{u}{G^{-1}\rho(H)} \Psi'\left(
                 \frac{u}{G^{-1}\rho(H)}\right) \right] \rd x\rd y
\nonumber\\[6pt]
	&=& -\int_{\mathfrak{D}} \Big(G^{-1}\rho(H)f,G^{-1}\rho(H)g\Big) \cdot
                     \nabla\left(\frac{u}{G^{-1}\rho(H)}
                        \right)
\nonumber\\
	&& \times \left[
                         \Psi\left(
                 \frac{u}{G^{-1}\rho(H)}\right)  +
		    \frac{u}{G^{-1}\rho(H)} \Psi'\left(
                 \frac{u}{G^{-1}}\rho(H)\right) \right] \rd x\rd y
\nonumber\\[6pt]
	&=& -\int_{\mathfrak{D}} \Big(G^{-1}\rho(H)f,G^{-1}\rho(H)g\Big) \cdot
                     \nabla \left[
		    \frac{u}{G^{-1}\rho(H)} \Psi\left(
                 \frac{u}{G^{-1}\rho(H)}\right) \right] \rd x\rd y
\nonumber\\[6pt]
	&=& -\int_{\mathfrak{D}}   \nabla\cdot\left\{ \Big( G^{-1}\rho(H)f, G^{-1}\rho(H)g\Big)
                    \left[
		    \frac{u}{G^{-1}\rho(H)} \Psi\left(
                 \frac{u}{G^{-1}\rho(H)}\right) \right]  \right\} \rd x\rd y
\nonumber\\[6pt]
	&=& \int_{\partial\mathfrak{D}}  \left\{ u(x,y,t)  \Psi\left(
                 \frac{u(x,y,t)}{G^{-1}(x,y)\rho(H)}\right) \big(f,g\big) \right\} \times (\rd x,\rd y).
\end{eqnarray}

\end{document}